\documentstyle[epsfig]{mn2e}
\begin{document}
\input BoxedEPS.tex
\SetEPSFDirectory{./}
\SetRokickiEPSFSpecial
\HideDisplacementBoxes

\title
[
The local sub-mm luminosity functions
]
{
The local sub-mm luminosity functions and predictions from Spitzer to
Herschel
}
\author[S. Serjeant \& D. Harrison]{
Stephen Serjeant$^{1}$ and Diana Harrison$^{2,3}$\\
$^1$ Centre for Astrophysics and Planetary Science, School of Physical
Sciences, 
University of Kent, Canterbury, Kent, CT2 7NZ, UK\\
$^2$ Astrophysics Group, Blackett Laboratory, Imperial College, 
Prince Consort Road,London SW7 2BW, UK\\
$^3$ Dept. of Physics, Cavendish Laboratory, Madingley Road, Cambridge
CB3 0HE, UK\\
}

\date{Accepted;
      Received;
      in original form 2003 March 4}
 
\pagerange{\pageref{firstpage}--\pageref{lastpage}}
\pubyear{2002}
\volume{}

\label{firstpage}

\maketitle


\begin{abstract} 
We present new
determinations of the local sub-mm luminosity functions, 
solving the ``sub-mm Olbers' Paradox.''
We also present 
predictions of source counts and luminosity functions in 
current and future far-infrared to sub-mm surveys.
Using the sub-mm colour temperature relations from the SCUBA Local
Universe
Galaxy Survey, and the discovery of excess $450\mu$m excess emission
in these galaxies, we interpolate and extrapolate the IRAS detections
to make predictions of the SEDs of all 15411 PSC-z galaxies 
from $50-1300\mu$m. Despite the long
extrapolations we find excellent agreement with (a) the $90\mu$m
luminosity function of Serjeant et al. (2001), (b) the $850\mu$m
luminosity function of Dunne et al. (2000), (c) the mm-wave photometry
of Andreani \& Franceschini (1996); (d) the asymptotic differential
and integral source count predictions at $50-1300\mu$m by
Rowan-Robinson (2001). 
We find the local $850\mu$m sub-mm luminosity density converges to 
$7.3\pm0.2\times10^{19}$ $h_{65}$ W Hz$^{-1}$ Mpc$^{-3}$. 
Remarkably, the local spectral luminosity density and
the extragalactic background light together strongly constrain the
cosmic star formation history for a wide class of evolutionary
assumptions. We find that the extragalactic background light, the
$850\mu$m $8$mJy source counts, and the $\Omega_\ast$ constraints all
independently point to a decline in the comoving star formation rate
at $z>1$. In order to reconcile this with direct determinations,
we suggest either there is 
a top-heavy initial mass function at high redshifts, and/or 
there is stronger evolution in the more 
luminous far-infrared galaxies than seen in the population as a whole. 
\end{abstract}

\maketitle

\section{Introduction}\label{sec:introduction}


The high expectations for sub-mm/mm-wave surveys have been fully and
variously realised. The discovery of a population of $z>1$ sub-mm
galaxies resolving up to $50\%$ of the extragalactic background light,
and the surprising lack of overlap with the hard X-ray Chandra
populations, have been among the key results in cosmology of the past
five years. 
The SCUBA survey of the Hubble Deep Field (HDF) demonstrated the
feasibility of blank-field surveys (Hughes, Serjeant, Dunlop,
Rowan-Robinson, et al. 1998, Serjeant et 
al. 2003), and several complementary sub-mm/mm-wave 
surveys have since been conducted
to a variety of depths and areal coverages (e.g. Scott et al. 2002;
Fox et al. 2002; Eales et al. 2000; Barger et al. 1999). 
These surveys
will be augmented with much  
larger surveys with (e.g.) 
SCUBA-2 (e.g. Holland et al. 2002), BOLOCAM on the CSO (e.g. Glenn et
al. 1998), the 
Balloon-borne Large Aperture Sub-mm 
Telescope (e.g. Tucker et al. 2004), the Large Millimeter Telescope
(e.g. Kaercher et al. 2000),
the Atacama Large 
Millimeter Array (ALMA), Planck High Frequency Instrument (HFI,
e.g. Lamarre et al. 2001), and
Herschel SPIRE (e.g. Griffin et al. 2001). 

It is not widely appreciated that the most reasonable prior
expectation for the SCUBA-HDF survey, based on the observed optical
HDF population, was {\it zero} sources at $850\mu$m: for
example, Thompson et al. (2001) model the near-IR/optical/near-UV SEDs
of the NICMOS-HDF galaxies, and derive sub-mJy/{\it micro}-Jy fluxes at
$850\mu$m. The mJy-level sub-mm galaxies found in blank-field surveys
on the whole can be reasonably regarded as a new galaxy population. 

One attractive interpretation of this new population is that it is
comprised at least in part of proto-spheroids (e.g. Dunlop 2001): the
star formation rates inferred
in sub-mm survey galaxies ($\stackrel{>}{_\sim}1000$ M$_\odot$/yr) are
sufficient to assemble a giant elliptical galaxy in $<1$Gyr. 
In support of this
interpretation, the K-band morphologies of at least some resemble
high-redshift radiogalaxies (Lutz et al. 2002), and provisional
measurements of the clustering are consistent with a high bias parameter
population at high redshift (Scott et al. 2002). Nevertheless, the
sub-mm galaxies 
are not unambiguously identified with proto-spheroids; 
Rowan-Robinson (2001), Efstathiou \& Rowan-Robinson (2002) and King \&
Rowan-Robinson (2003) instead
postulate that the blank-field sub-mm survey galaxies are dominated
in the sub-mm by cirrus heated by their interstellar radiation
fields, rather than ultraluminous star formation. In
support of this interpretation, Efstathiou \& Rowan-Robinson (2002) 
model the
SEDs of sub-mm galaxies and find them consistent with their cirrus
radiative transfer models. The existence of cool colour temperature
components in local galaxies, physically distinct from the warmer dust
dominating the $60\mu$m emission, have been pointed out by
many authors (e.g. Dunne et al. 2000 and refs. therein, Stickel et
al. 2000,  Siebenmorgen et al. 1999, Domingue et al. 1999,
Trewhella et al. 2000; Haas et al. 2000.) 
If this cool
component dominates the observed sub-mm fluxes of high-$z$ blank field
sub-mm/mm-wave survey sources, then the emission should be extended
over $\sim 1''$ rather than confined to a compact $\sim 0.1''$
ultraluminous starburst (Efstathiou \& Rowan-Robinson 2002, Lee \&
Rowan-Robinson 2003, Kaviani et al. 2002). This distinction will be
possible with the planned ALMA. 

In the meantime, 
the evolution of the sub-mm galaxy population can be strongly constrained
by the integrated extragalactic background light, the local
multiwavelength luminosity functions, and the source counts. 
%
The local $850\mu$m luminosity
function was derived in  
the SCUBA Local Universe Galaxy Survey (SLUGS, Dunne et al. 2000) from
their SCUBA photometry of the IRAS Bright Galaxy Survey. A curious
aspect of their luminosity function was that the faint end slope was
not sufficiently shallow for the local luminosity density to converge,
which the authors referred to as the ``sub-mm Olbers' paradox''. This
is a pity from the point of view of modelling the high redshift
population, since the integrated extragalactic background light is a
key constraint. 
In order
to find the expected flattening of the luminosity function slope at
lower luminosities, the SLUGS 
survey is currently being extended with SCUBA photometry of
optically-selected galaxies. Meanwhile, several authors have
attempted to use the colour temperature -- luminosity relation found
in SLUGS to transform the $60\mu$m luminosity function to other
wavelengths, and hence constrain the high-redshift evolution
(e.g. Lagache, Dole \& Puget 2003, Lee \& Rowan-Robinson 2003, Chapman
et al. 2003). The 
discovery of an
additional excess component at $450\mu$m (Dunne \& Eales 2001)
relative to their initial colour temperature -- luminosity relation
is only rarely included in such models (e.g. Lagache, Dole \& Puget
2003), and there are problems in accurately representing
the population mix of galaxies. 

In this paper we take an
alternative approach to determining the multiwavelength local
luminosity functions. 
We model the spectral energy distributions (SEDs) of all 15411 
PSC-z galaxies (Saunders et al. 2000), constrained by all available
far-infrared and sub-mm colour-colour relations from SLUGS and
elsewhere.
This guarantees the correct local population mix at every wavelength
and minimises the assumptions about the trends of SED shape with
luminosity (see e.g. Rowan-Robinson et al. 2001, in which the
differences in assumed population mix result in rather different
conclusions to this work). 
Our phenomenological approach is sufficient to determine (for
example) the $850\mu$m luminosities to within a factor of $2$ in
individual galaxies. While
this is not accurate enough to serve as calibrations for sub-mm
missions, except statistically over the whole population, 
it is more than sufficiently accurate 
for determining local luminosity functions, which we present here from
far-infrared to millimetric wavelengths.  
We use these local luminosity functions to model the
high-$z$ evolution, constrained by the extragalactic background light
spectrum, the total cosmological matter density in stars $\Omega_\ast$,  
and the sub-mm source counts. 

Our SED models are accurate enough to identify potential calibrators
for further photometric study. The SEDs 
can also be used to determine the bright source
contributions to current and future far-infrared and sub-mm/mm-wave
surveys. 
Predictions for bright source catalogues are also 
essential tools for the
planning and execution of future wide area far-infrared missions. 
For example, the catalogues are useful for 
both tertiary flux calibration and astrometric calibration
in all-sky 
surveys (e.g. Planck HFI, Astro-F, e.g. Pearson et al. 2001, 
Shibai 2002), and also 
for mission strategy (such as deep survey fields avoiding 
bright sources). 

Section
\ref{sec:method} describes our modelling of the spectral energy
distributions, including observational tests of our methodology. 
Section \ref{sec:results} presents the results, such as
multiwavelength local luminosity functions, and bright source counts
for Planck HFI, Herschel SPIRE, and Astro-F
The
implications 
of our results are discussed in section \ref{sec:discussion}. 

Throughout this paper we adopt the ``concordance'' cosmology 
of $\Omega_{\rm M}=0.3$, $\Omega_\Lambda=0.7$ and $H_0=65$ km s$^{-1}$
Mpc$^{-1}$. 

\section{Method}\label{sec:method}
\subsection{$850\mu$m predictions}

The PSC-z survey is a redshift survey of IRAS
galaxies, to a depth of $0.6$Jy at $60\mu$m. PSC-z covers $84\%$ of
the sky, and has a median redshift of $8400$ km s$^{-1}$. Further
details can be found in Saunders et al. (2000). 

\begin{figure}
  \ForceWidth{3.0in}
  \BoxedEPSF{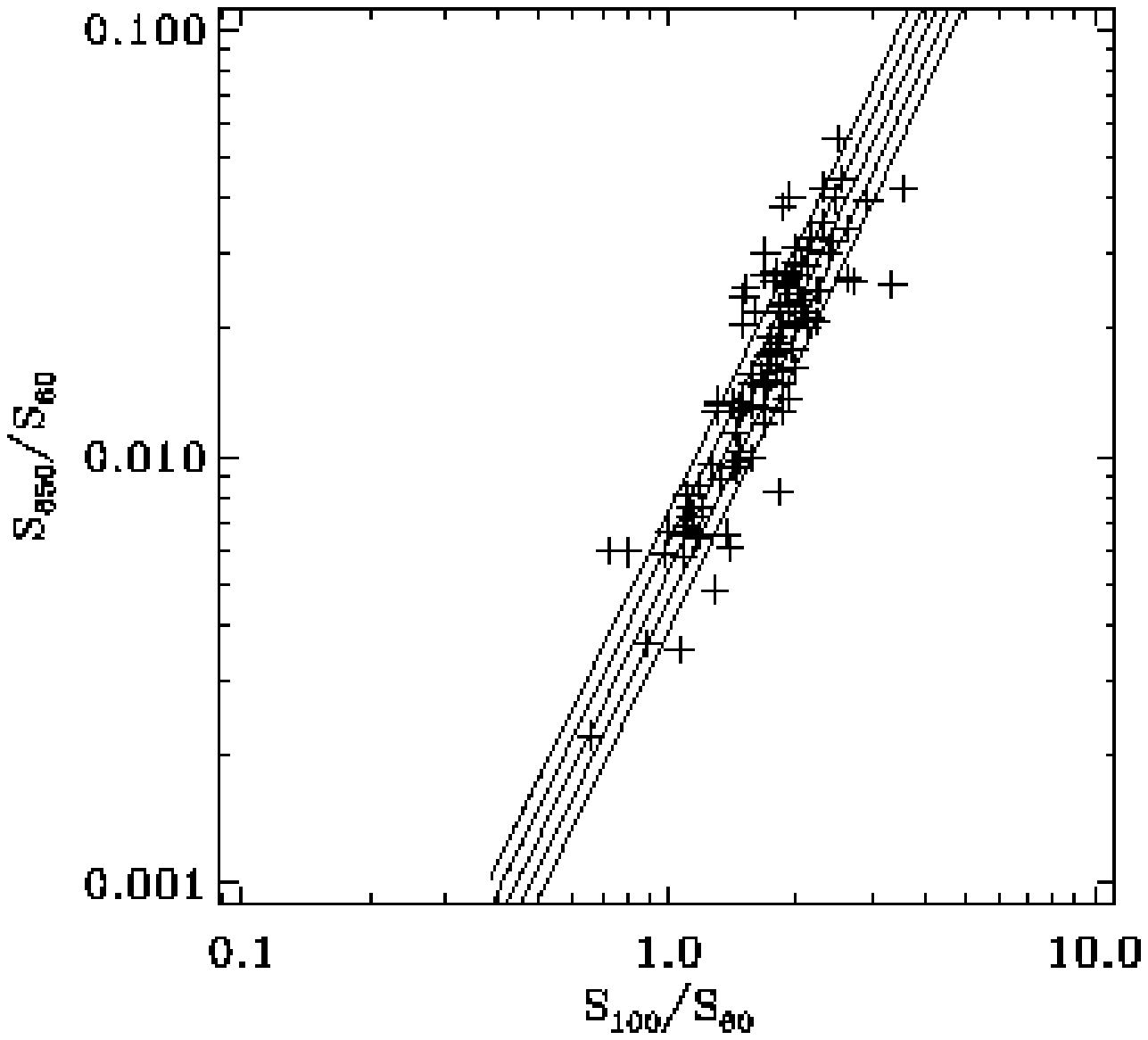}
\caption{\label{fig:colour-colour}
Sub-mm:far-infrared colour-colour plane for the Dunne et al. (2000)
sample. From top to bottom, lines show the predictions for grey body
indices of $\beta=1.1$, $1.2$, $1.3$, $1.4$, $1.5$. 
}
\end{figure}

Our first stage in modelling the far-infrared to mm-wave SEDs of
IRAS PSC-Z galaxies is to determine their $850\mu$m fluxes. To do this 
we use the results from the $850\mu$m photometry programme of the IRAS 
Bright Galaxy Sample by Dunne et al. (2000). These authors found a
strong correlation between far-infrared luminosity and
sub-mm:far-infrared colour.  
The far-infrared luminosity correlates strongly with the $60-100\mu$m
colour, so an alternative approach to modelling the luminosity-colour
relation is to fit to the $100-60\mu$m vs. $850-60\mu$m colour-colour
plane. This is plotted in figure 1, together with the predictions from 
grey-body emission:
\begin{equation}\label{eqn:greybody}
S_\nu = \nu^\beta B_\nu 
     = \nu^\beta \frac{2h\nu^3}{c^2} \frac{1}{\exp(h\nu/(kT)) - 1}
\end{equation}
The lines in figure \ref{fig:colour-colour} correspond to
$\beta=1.1,1.2,1.3,1.4,1.5$. Dunne 
et al. (2000) also modelled the colour temperatures and $\beta$ values 
for all their sample using limited $450\mu$m data, and found a median
$\beta=1.3$ which is in excellent agreement with the best fit to figure 
\ref{fig:colour-colour}. Physically, these galaxies are very unlikely
to be single 
temperature or even dual 
temperature systems (e.g. Efstathiou, Rowan-Robinson \& Siebenmorgen
2000) but all that matters for our current purpose is that we have a
phenomenological model which can be used to interpolate to other
frequencies. 

We use the co-added or extended ADDSCAN IRAS fluxes (Saunders et
al. 2000). 
For the $29\%$ of 
sources with only $100\mu$m limits, we use the non-linear 
$60\mu$m-$100\mu$m luminosity-luminosity correlation (i.e., the 
well-known trend of IRAS
colour temperature correlating with luminosity) to estimate the
$100\mu$m fluxes. Redshift information is available for almost all of
PSC-Z, but because redshift is degenerate with colour temperature
this information is not necessary for sub-mm flux prediction where
$100\mu$m detections are available. 

\subsection{$450\mu$m predictions and interpolations/extrapolations}
While the phenomenological model above is a good predictor of the
$850\mu$m fluxes of IRAS galaxies, there is an additional component at 
shorter wavelengths discovered by Dunne \& Eales (2001). 

The combined $850\mu$m, $450\mu$m, $100\mu$m and $60\mu$m data of the
IRAS Bright Galaxy Survey led Dunne \& Eales (2001) to argue for
two $\beta=2$ grey-body components, because the excess flux in the
short-wavelength sub-mm data could not be fit by any  
single grey-body template. As the two component model is a
four-parameter fit (two temperatures and two normalisations) with the
emissivity fixed,
four-band photometry is just sufficient in principle 
to define the fit uniquely. 

Together with the IRAS measurements, our prescription for $850\mu$m
fluxes gives us a total of three bands on which to base SED
interpolations. We can obtain a further phenomenological prediction
for the $450\mu$m fluxes on the linear 
$60:850\mu$m vs. $60:450\mu$m colour-colour correlation, for which the
IRAS Bright Galaxy Survey yields 
\begin{equation}\label{eqn:450}
\log_{10}(S_{60}/S_{450}) = \log_{10}(S_{60}/S_{850}) - 0.924
\end{equation}
These prescriptions should be sufficient to define the sub-mm fluxes
to within a factor of two. 

We then numerically solve for the parameters of the two temperature
components, using the 
four-band measured and predicted photometry. Note that this two
component fit uses a grey body index of $\beta=2$, which is
not consistent with the $\beta=1.3$ model used to predict the
$850\mu$m fluxes. This does not affect the self-consistency or utility 
of the two-component fit, because the only relevant information 
is that a numerical scheme of whatever nature is available to predict
the $850\mu$m fluxes of the IRAS PSC-Z galaxies. Also, it is worth
stressing again that more physically reasonable models incorporate a
continuum of temperature components; all that matters for our current
purpose is to obtain a phenomenological parametric model on which to
base predictions of the far-IR to mm-wave SEDs. In the following
section we will test the consistency of the phenomenological
predictions using a variety of observational constraints. 

The dual-temperature model incorporates the available photometric 
constraints on
the far-infrared to mm-wave SEDs of IR-luminous galaxies, and provides 
the best available interpolations and extrapolations from
$50-1300\mu$m in an all-sky catalogue. The PSC-Z predictions for
Astro-F, Herschel, Planck HFI and ISO are available 
as an IDL database from the authors. Also available is interpolation
software to provide further predictions at further arbitrary
wavelengths. 

%
%

\section{Results}\label{sec:results}

\subsection{Observational tests}
In figure 2 we plot the observed $1.25$mm photometry of Andreani \& 
Franceschini (1996), compared with our predictions. These observations are
subject to a systematic uncertainty of up to a factor of $2$ due to
aperture corrections. Not withstanding a large systematic shift in flux
calibration, our predictions are in good agreement with these
observations. 

\begin{figure}
  \ForceWidth{4.0in}
  \hSlide{-1cm}
  \BoxedEPSF{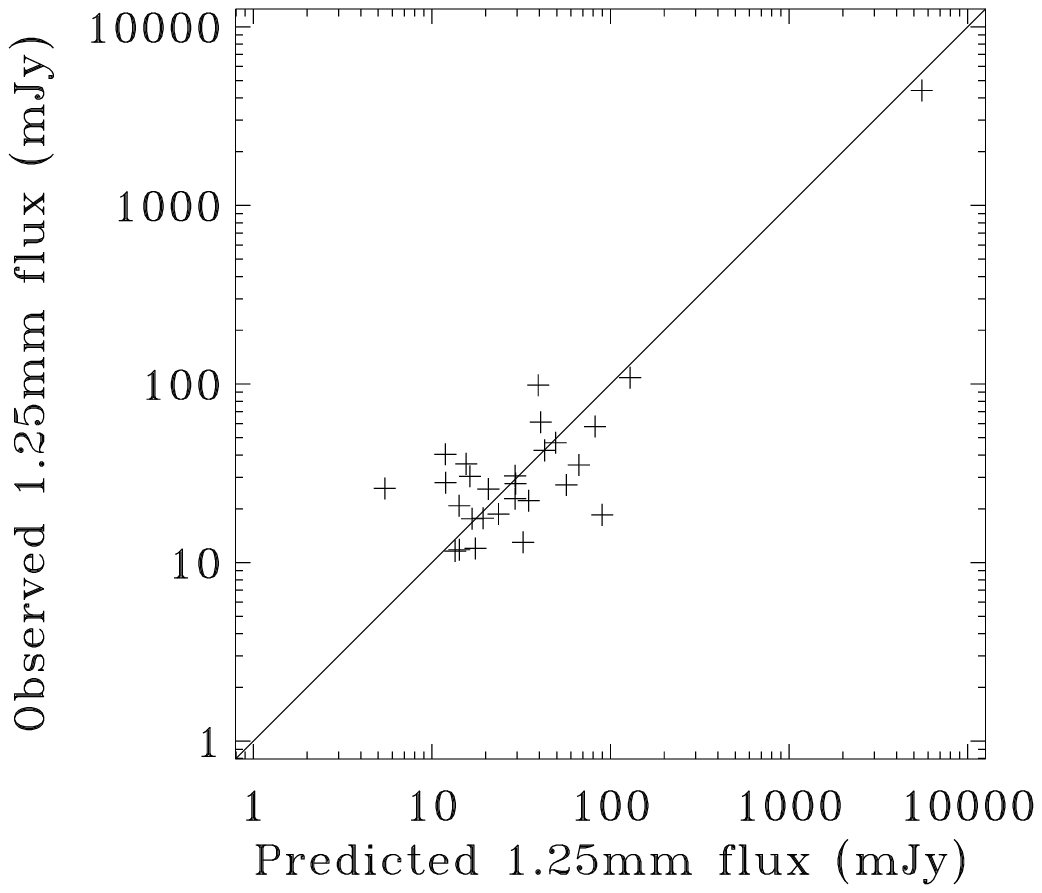}
\caption{\label{fig:franceschini}
Comparison of predicted and observed $1.25$mm fluxes for the sample of 
Andreani \& Franceschini 1996, using their $\alpha_{\rm
mm}=\alpha_{\rm o}/3$ aperture correction. Their alternative 
$\alpha_{\rm mm}=\alpha_{\rm o}$ aperture corrections result in an
upward revision of the observed fluxes by a factor of
$\sim\times1.5-2$. The straight line indicates the locus of exact
agreement between observations and predictions.  
}
\end{figure}

\begin{figure}
  \ForceWidth{4.0in}
  \hSlide{-1cm}
  \BoxedEPSF{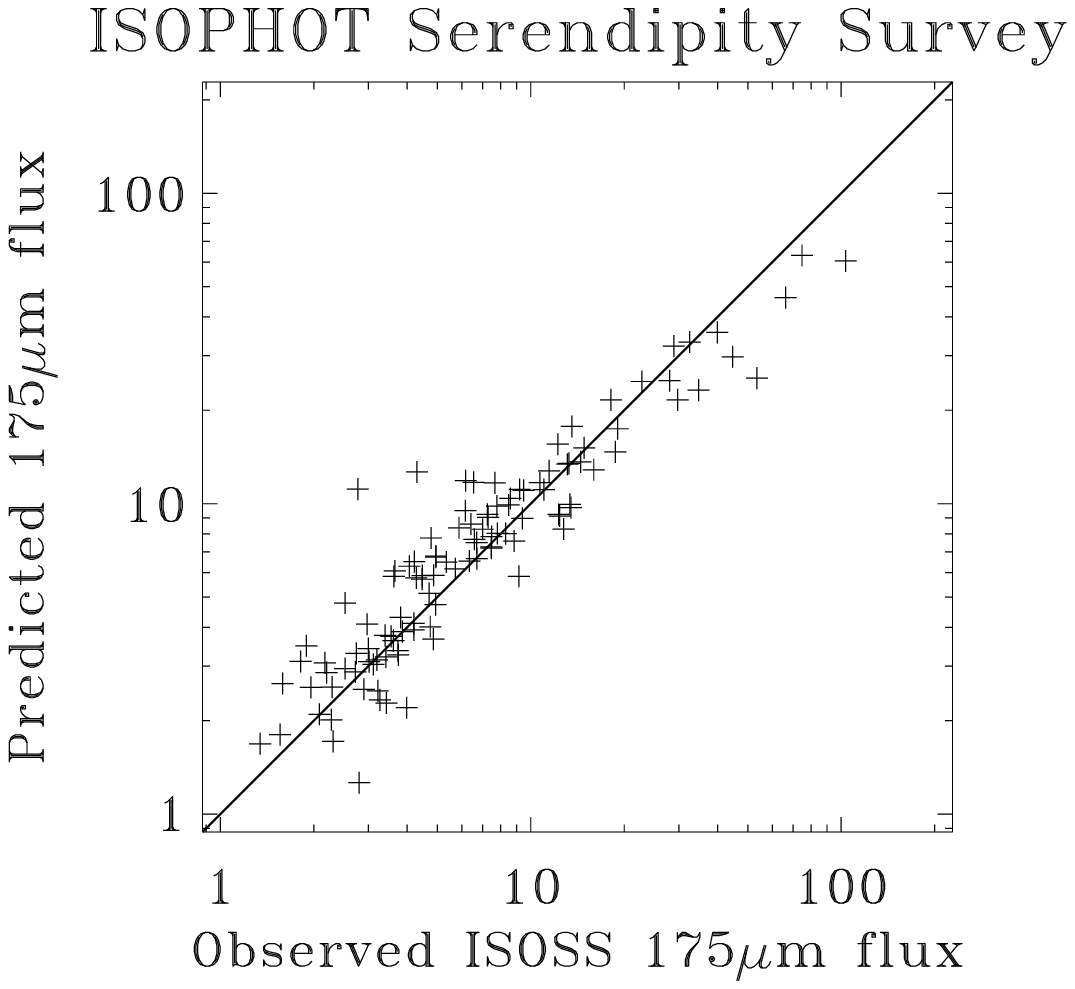}
\caption{\label{fig:stickel}
Comparison of predicted and observed $175\mu$m fluxes for the ISOPHOT
Serendipity Survey of Stickel et al. (2000), using the best
available current flux calibration (Stickel. priv. comm.). 
}
\end{figure}

A stronger tests of our methodology comes from the Serendipity Survey
of the Infrared Space Observatory (ISOSS), taken at $175\mu$m by the
ISOPHOT instrument (e.g. Stickel et al. 2000). The survey covered
$\sim10\%$ of the sky, and although the completeness is not yet well
understood the flux calibration is well-determined enough to provide a
useful test. Figure
\ref{fig:stickel} shows the predicted $175\mu$m fluxes against those
observed. The agreement is excellent. This also demonstrates that the
cool dust excess reported by Stickel et al. (2000) is identical to that
reported  by Dunne \& Eales (2001), at least insofar as fitting of
SEDs with colour temperature components is concerned. (This does not
necessarily imply the $175\mu$m-emitting dust is cospatial to the
$450\mu$m-emitting dust, e.g. section \ref{sec:luminosity_density}.)

\begin{figure}
  \ForceWidth{4.0in}
  \hSlide{-1cm}
  \BoxedEPSF{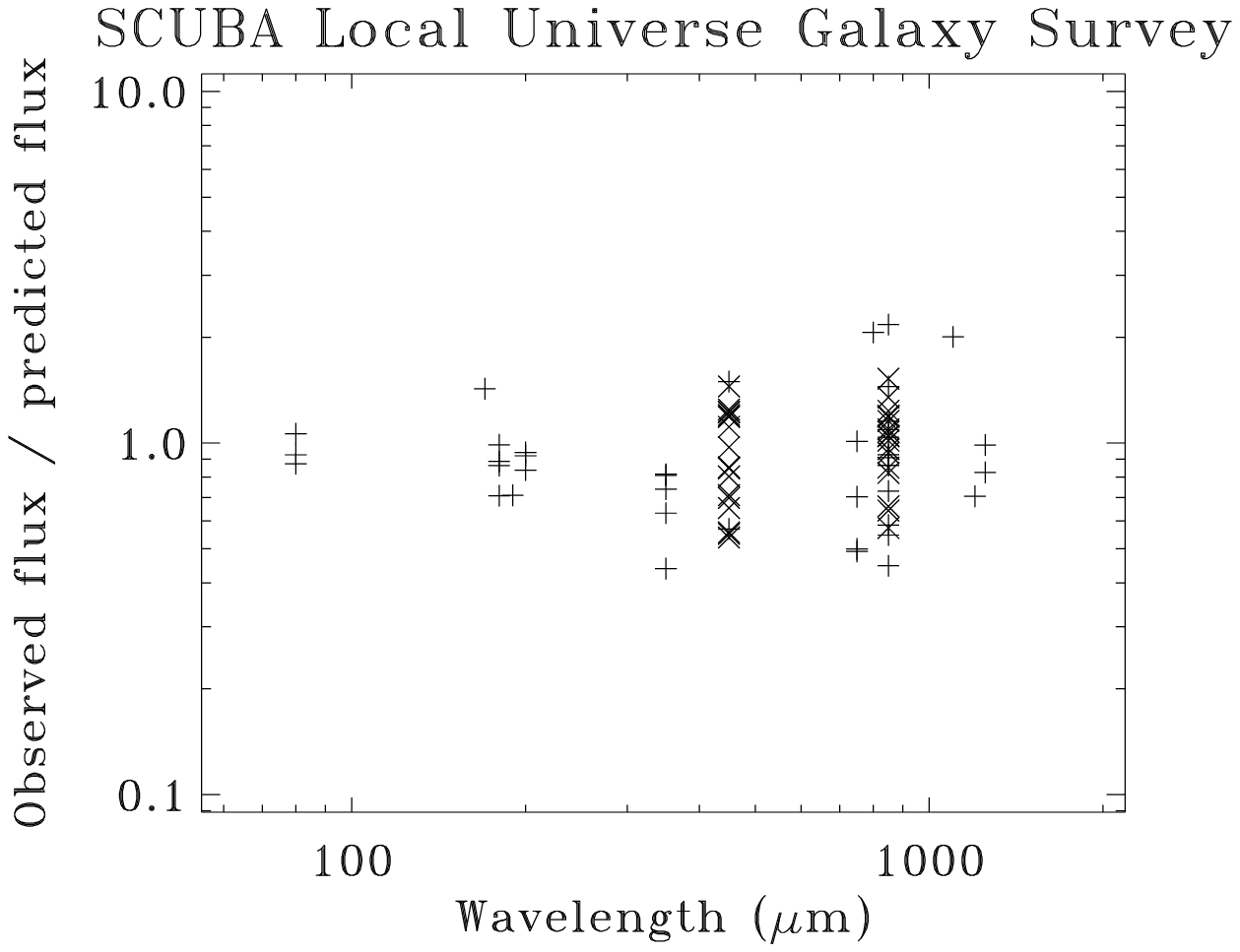}
\caption{\label{fig:slugs_multiwavelength}
Comparison of the observed SEDs of SLUGS galaxies with the predictions
of this paper. The $\times$ symbols indicate the SLUGS measurements
themselves, while the $+$ symbols represent the supplementary data
compiled from the literature by Dunne \& Eales (2001). 
}
\end{figure}

There are few other examples of objects with sufficient multi-wavelength
coverage to provide tests of our methodology (a problem also
encountered by other authors: e.g. Hughes et
al. 2002). In figure \ref{fig:slugs_multiwavelength} we plot the ratio of
our predictions with the observations for the SLUGS survey, as a
function of wavelength. The dispersion is a measure of the error in
our predictions for individual objects, while the mean values 
demonstrate that the systematic errors in our methodology are
typically only at the few percent level in the wavelength range
covered. Our systematic agreement is slightly better than that of Dale
et al. (2001, 2002) who use a single-component SED but with a varying
grey body index. 

\begin{figure}
  \ForceWidth{4.0in}
  \hSlide{-1cm}
  \BoxedEPSF{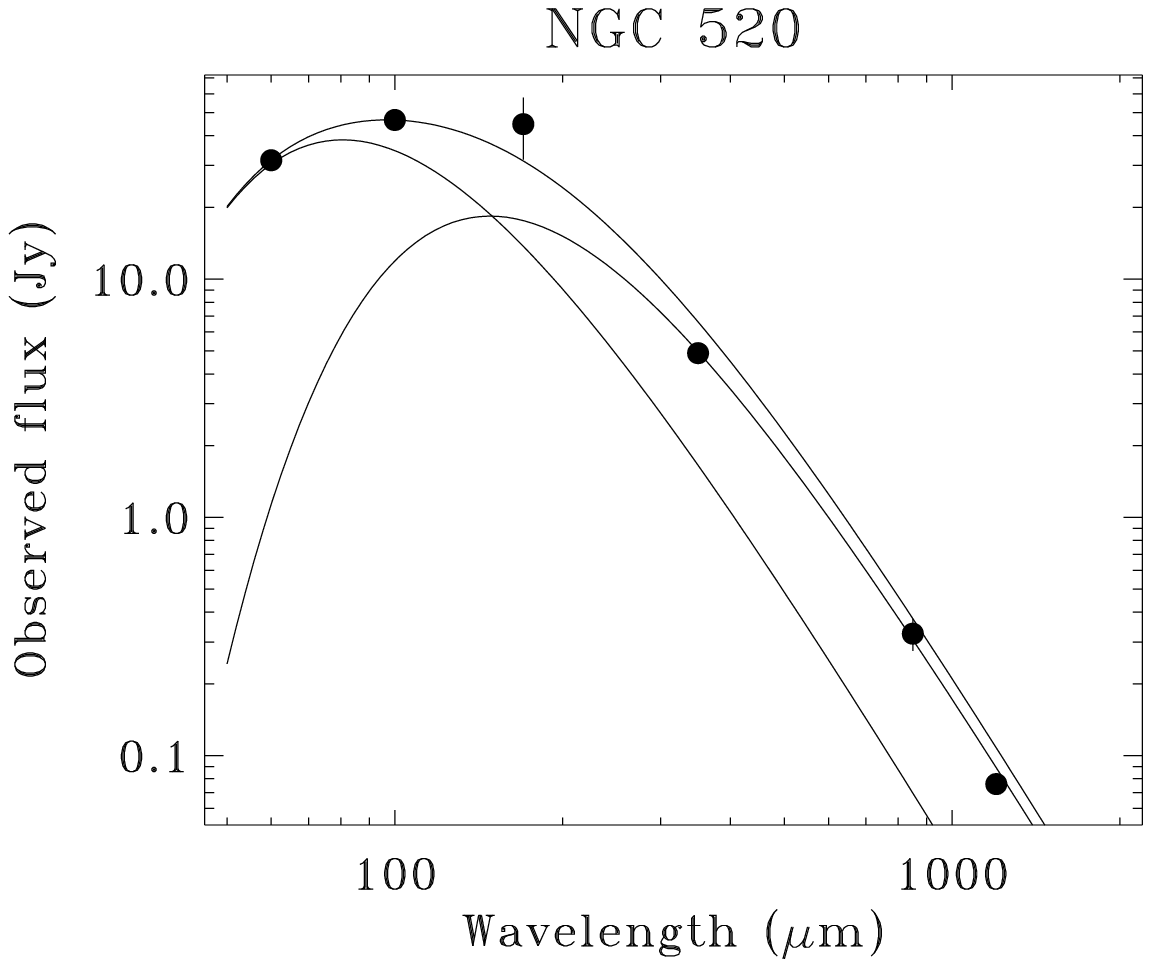}
\caption{\label{fig:ngc520}
Comparison of the observed SED of NGC520 with the predictions. The
cool and warm components are shown separately, and the sum (also
shown) is in good agreement with the observations. References for the
photometry are in Dunne \& Eales (2001). 
}
\end{figure}

One of the individual galaxies in the SLUGS survey, also covered by
the ISOSS survey, is shown in figure \ref{fig:ngc520}. The agreement
is again very good. Note that this is {\it not} a fit to the
multi-wavelength data; instead, the model is the mean SED for a galaxy
with the $60:100\mu$m colour of this galaxy. 

\begin{figure}
  \ForceWidth{4.0in}
  \hSlide{-1cm}
  \BoxedEPSF{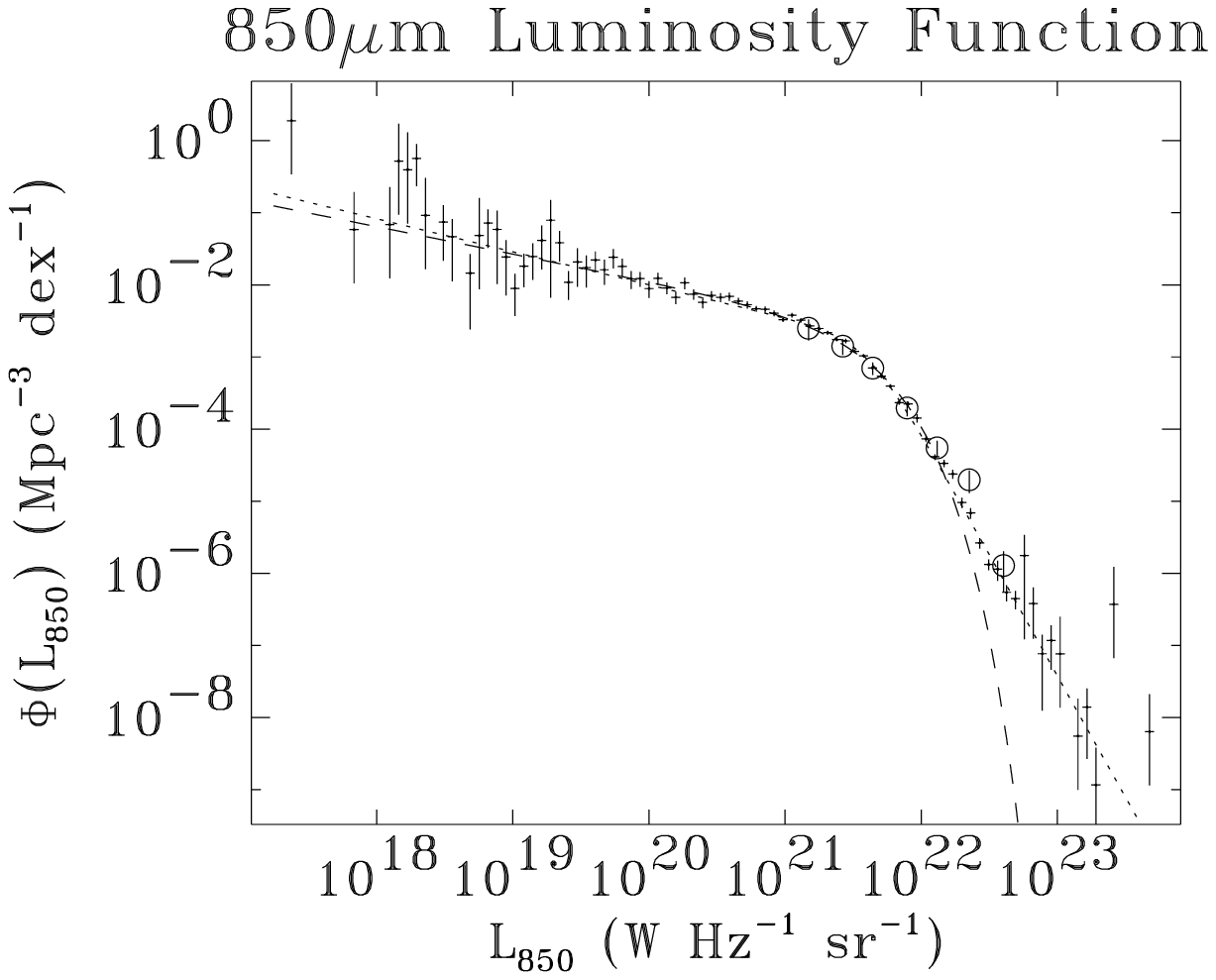}
\caption{\label{fig:850_lf}
Projected local $850\mu$m luminosity function from PSC-Z (error bars)
assuming pure luminosity
evolution of $(1+z)^3$,
compared with the directly measured luminosity function from Dunne et
al. 2000 (plotted as open circles). Also plotted are the best fit
Schechter and double power 
law functions (dashed and dotted lines respectively). 
}
\end{figure}

\subsection{Multi-wavelength local luminosity
functions}\label{sec:luminosity_functions}  

Our sample is $60\mu$m-selected, but that does not preclude us from
constructing a luminosity function at another wavelength
provided there are no large
populations that would be missing from our sample at {\it any}
redshift, compared to a flux limited sample at that other
wavelength. Serjeant et al. (2001) describe the methodology for
deriving luminosity functions in the face of complicated selection
functions, and the conditions for the validity of such derivations. 
In summary, the number density in a given luminosity bin is given by
\begin{equation}
\Phi=\Sigma(V_{{\rm max},i})^{-1}
\end{equation}
and the error on this quantity is given by 
\begin{equation}
\Delta\Phi=\sqrt{\Sigma(V_{{\rm max},i})^{-2}}
\end{equation}
Here, $V_{{\rm max},i}=V(z_{{\rm max},i})$ where $V$ is the comoving
volume, and $z_{{\rm max},i}$ is the redshift at which object $i$
would be observed at the $60\mu$m PSC-z flux limit, if all other 
intrinsic properties of that object were kept the same. 
Our methodology differs slightly from that of Takeuchi et al. (2003),
in that those authors used different K-corrections and a different
sub-samples (warm and cool) of the PSC-z survey, as well as making a
correction for density fluctuations. These differences in methodology 
lead to small changes in (for example) the faint end slope, but are
too small to affect the conclusions of our paper.  

\begin{figure}
  \ForceWidth{4.0in}
  \hSlide{-1cm}
  \BoxedEPSF{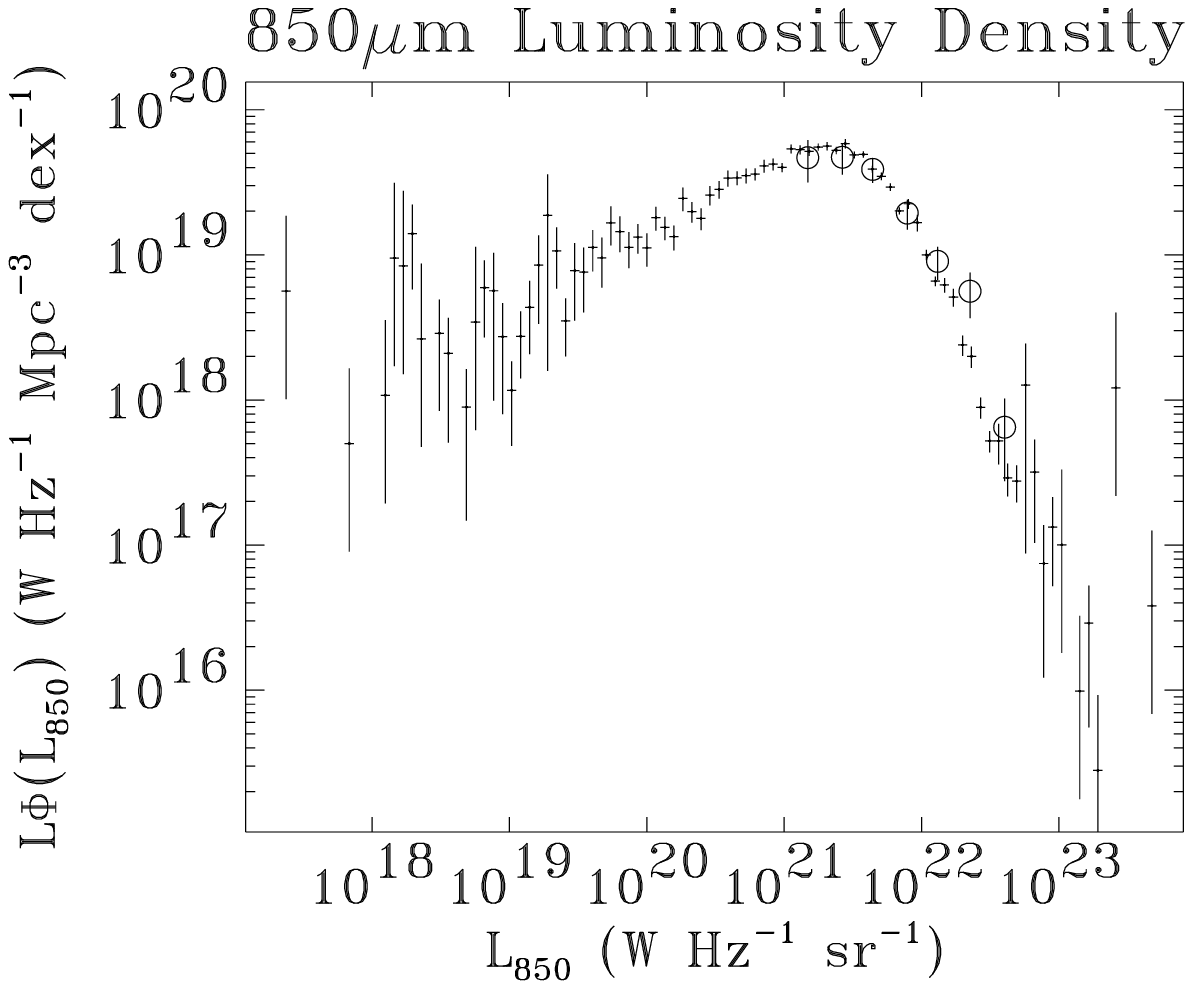}
\caption{\label{fig:850lumdens}
Projected local $850\mu$m luminosity density function from PSC-Z
(error bars), compared with the directly measured luminosity density
from Dunne et al. 2000 (plotted as open circles). This is derived
directly from figure \ref{fig:850_lf} after removal of the sr$^{-1}$
factor in the luminosities. Note that the
addition of the PSC-z data adds the first evidence of convergence in
the local luminosity density at $850\mu$m. 
}
\end{figure}

Although we have found that the $850\mu$m predictions for individual
galaxies are only good to within a factor of $\sim2$ (figure
\ref{fig:slugs_multiwavelength}), this is nevertheless accurate enough
for the 
determination of the local luminosity function. Note that
these luminosities are estimated directly from the colour-colour
relations, and are not interpolated using our two-component dust
models (except for the K-correction terms). 
In figure \ref{fig:850_lf}, we plot the $1/V_{\rm max}$ 
$850\mu$m luminosity function for 
PSC-Z assuming $(1+z)^3$ pure luminosity evolution (an assumption
which makes a small correction to only the brightest data points), and
compare it to 
the observed luminosity function from Dunne 
et al. (2000). 
The agreement
is excellent, and as with further multi-wavelength luminosity
functions below is sufficient to determine the convergence of the
local luminosity density. This is shown in the case of the $850\mu$m
luminosity density in figure \ref{fig:850lumdens}, where the PSC-z
constraints are compared to the existing SLUGS data. The SLUGS
data reaches the peak of the luminosity density, but would need to
extend around $\sim\times10$ fainter to determine the convergence in
the background. 
Using PSC-z, we find that local luminosity
density at $850\mu$m converges at $7.3\pm0.2\times10^{19}$
$h_{65}$ W Hz$^{-1}$ Mpc$^{-3}$, solving the sub-mm Olbers' 
Paradox. The constraints at other wavelengths 
will be discussed in further detail in
section \ref{sec:luminosity_density}. 

\begin{figure}
  \ForceWidth{4.0in}
  \hSlide{-1cm}
  \BoxedEPSF{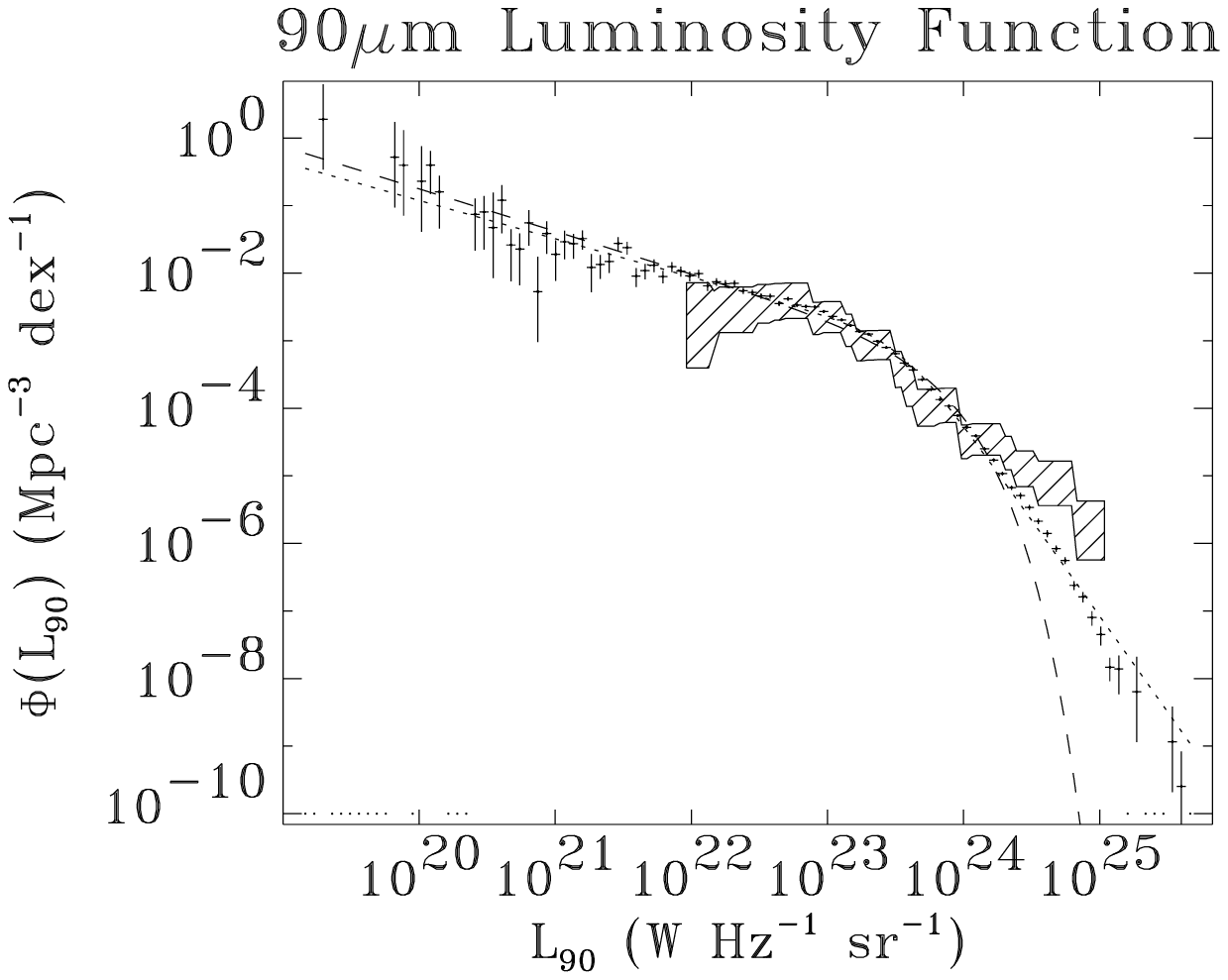}
\caption{\label{fig:90_lf}
Projected local $90\mu$m luminosity function from PSC-Z (error bars),
compared with the directly 
measured luminosity function from ELAIS (shaded
area) from Serjeant et al. 2001. In both cases pure luminosity
evolution of $(1+z)^3$ was assumed. Also plotted are the best fit
Schechter and double power 
law functions (dashed and dotted lines respectively). 
}
\end{figure}

\begin{figure*}
\vspace*{-0.5cm}
\ForceWidth{3.5in}\hSlide{-3cm}\BoxedEPSF{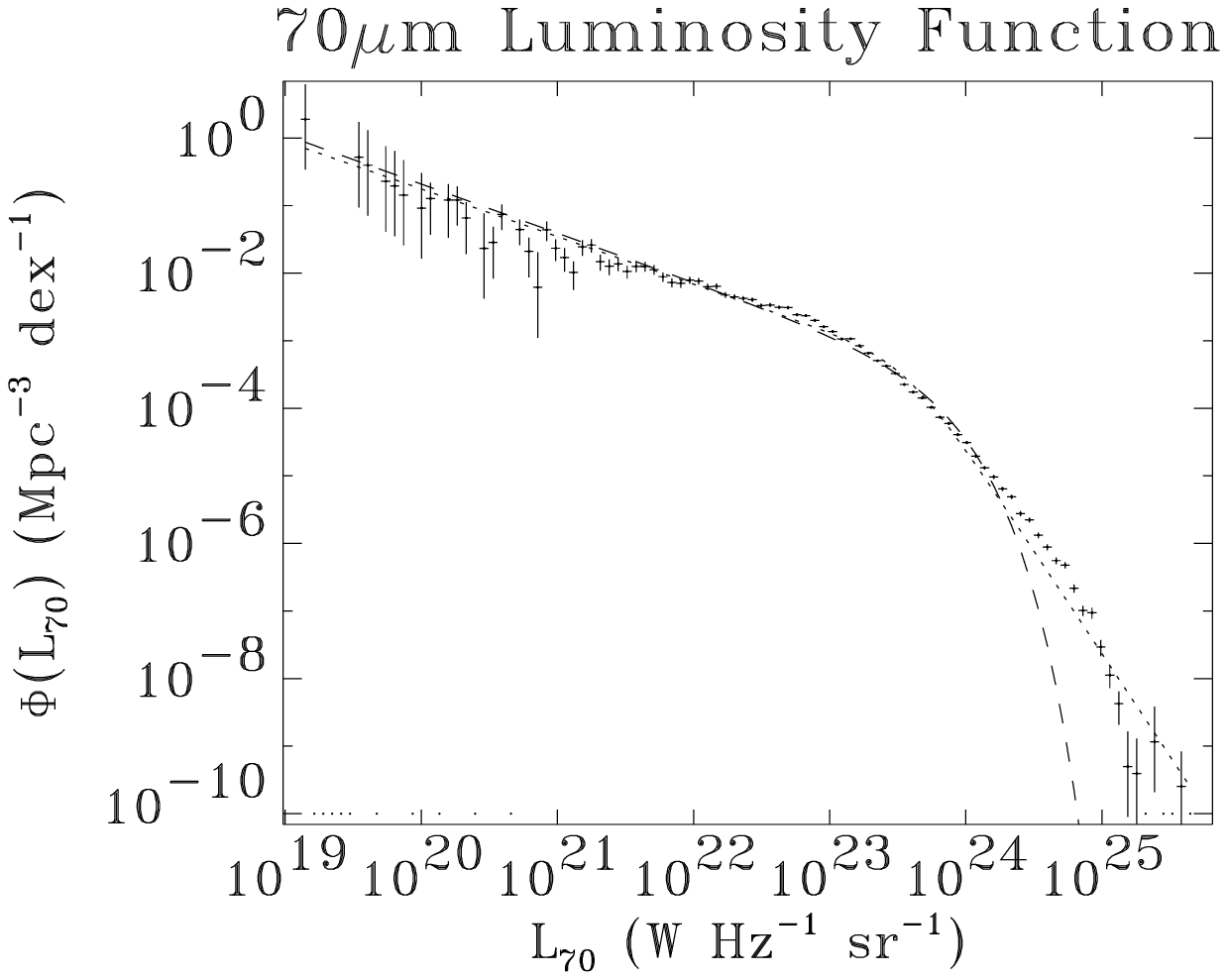}
\ForceWidth{3.5in}\hSlide{5cm}\vspace*{-6.4cm}\BoxedEPSF{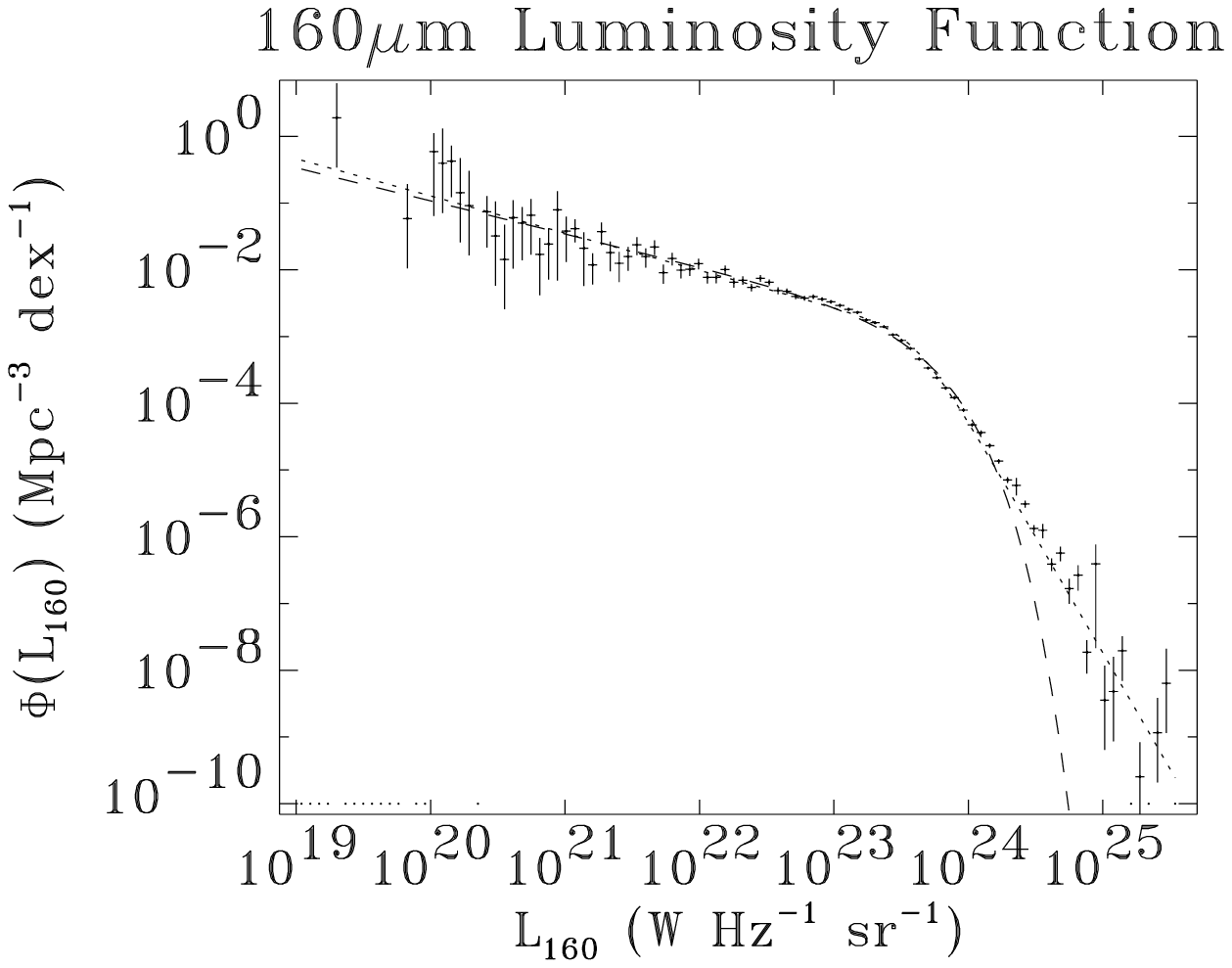}
\vspace*{-1cm}
\ForceWidth{3.5in}\hSlide{-3cm}\BoxedEPSF{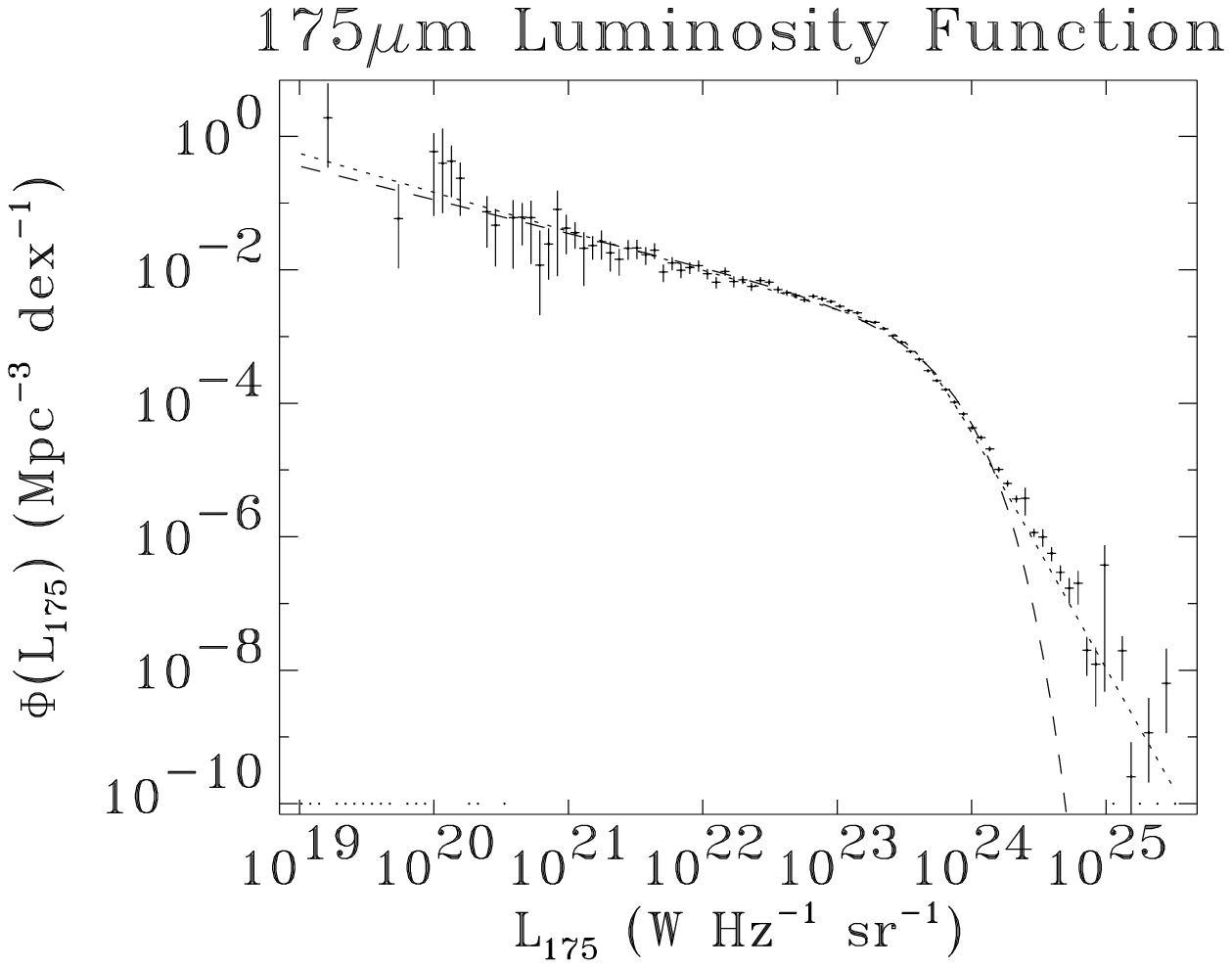}
\ForceWidth{3.5in}\hSlide{5cm}\vspace*{-6.4cm}\BoxedEPSF{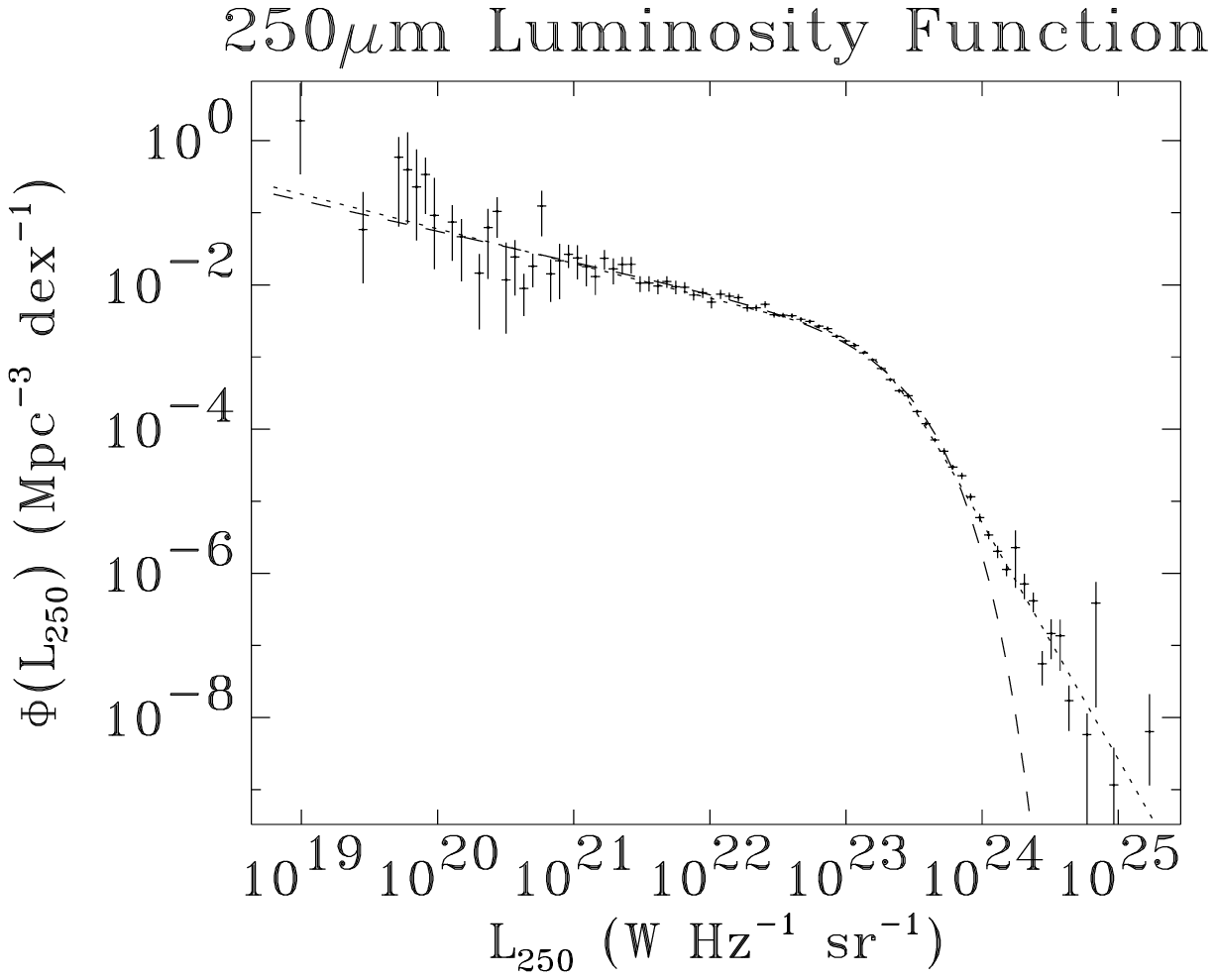}
\vspace*{-1cm}
\ForceWidth{3.5in}\hSlide{-3cm}\BoxedEPSF{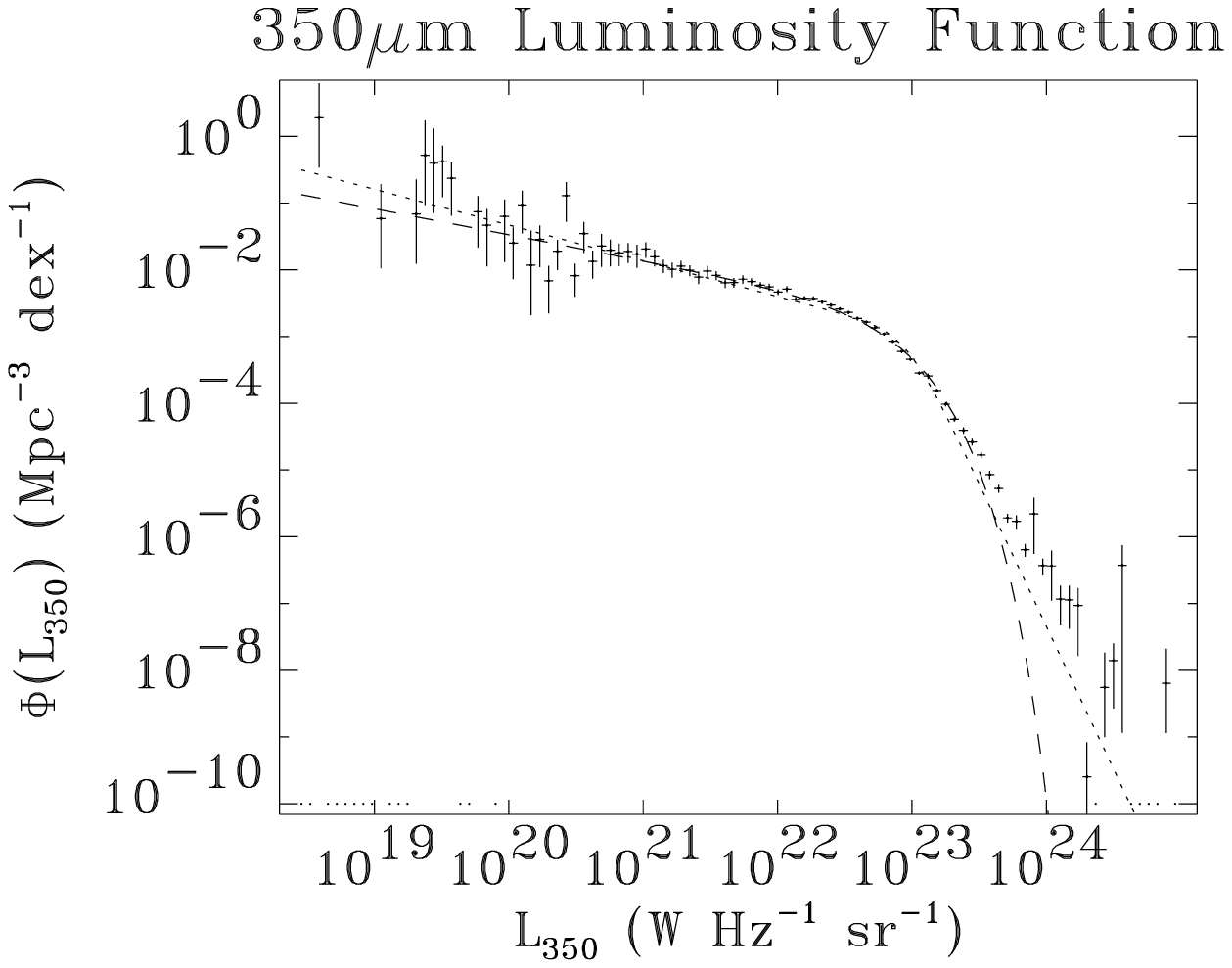}
\ForceWidth{3.5in}\hSlide{5cm}\vspace*{-6.4cm}\BoxedEPSF{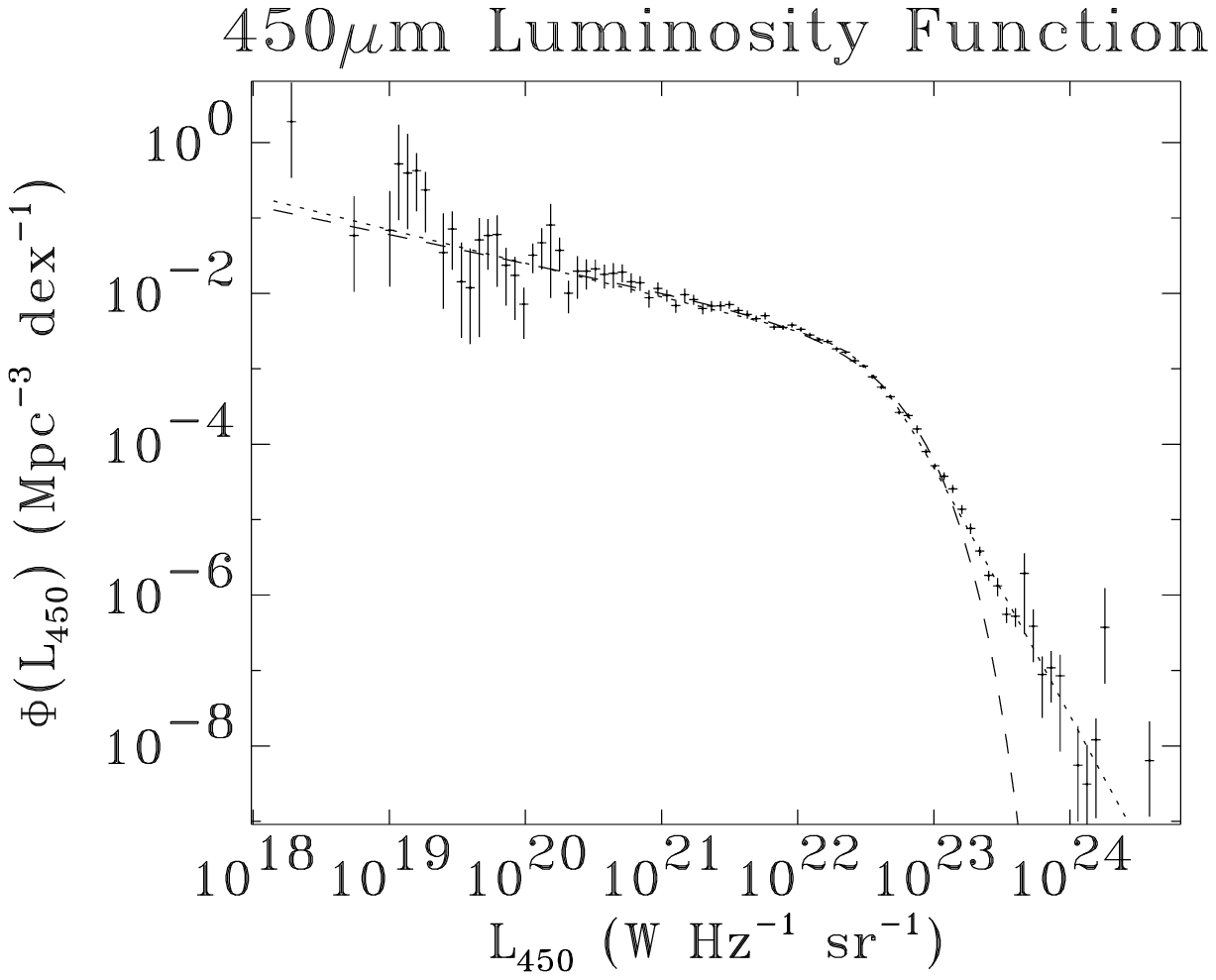}
\vspace*{-1cm}
\ForceWidth{3.5in}\hSlide{-3cm}\BoxedEPSF{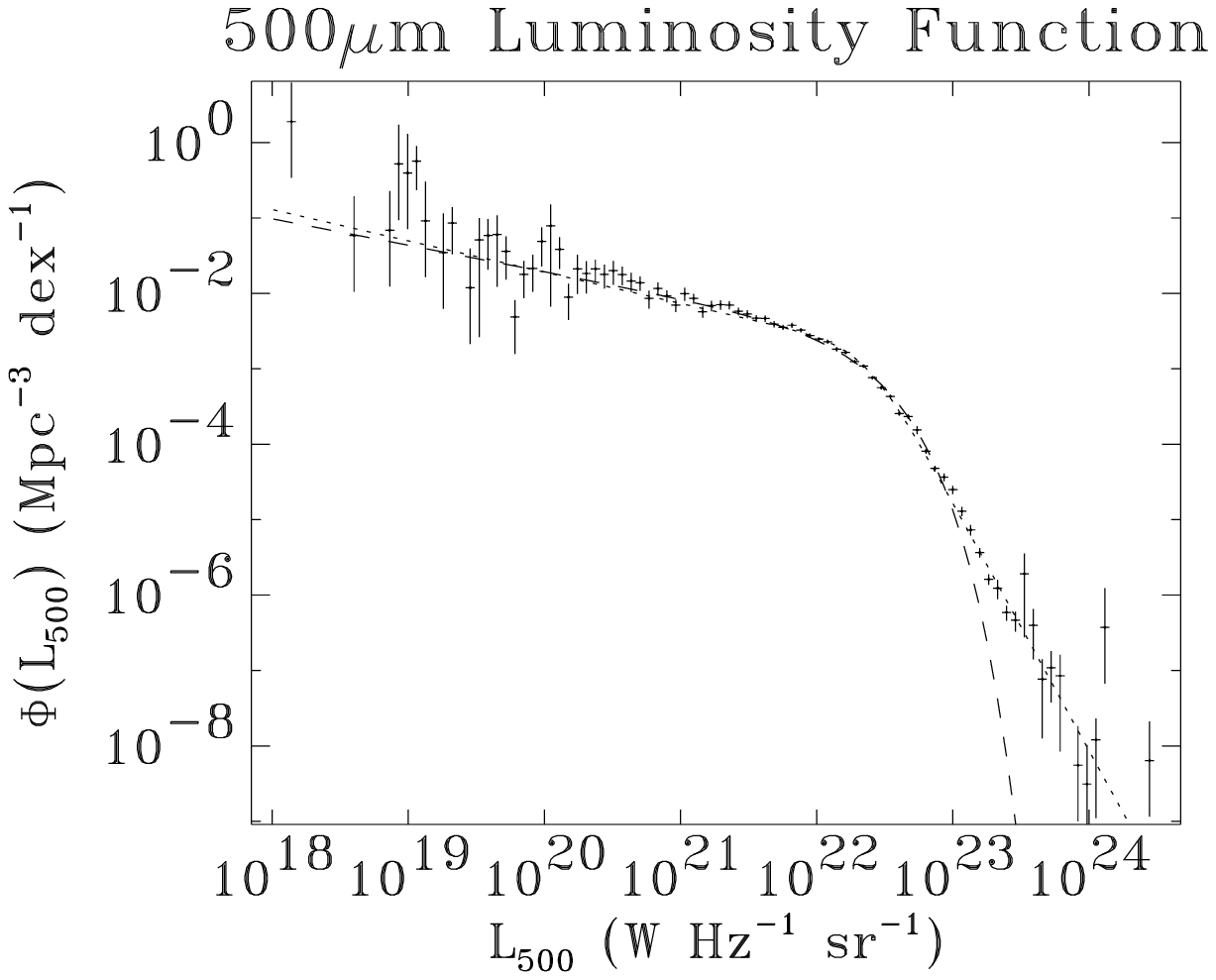}
\ForceWidth{3.5in}\hSlide{5cm}\vspace*{-6.4cm}\BoxedEPSF{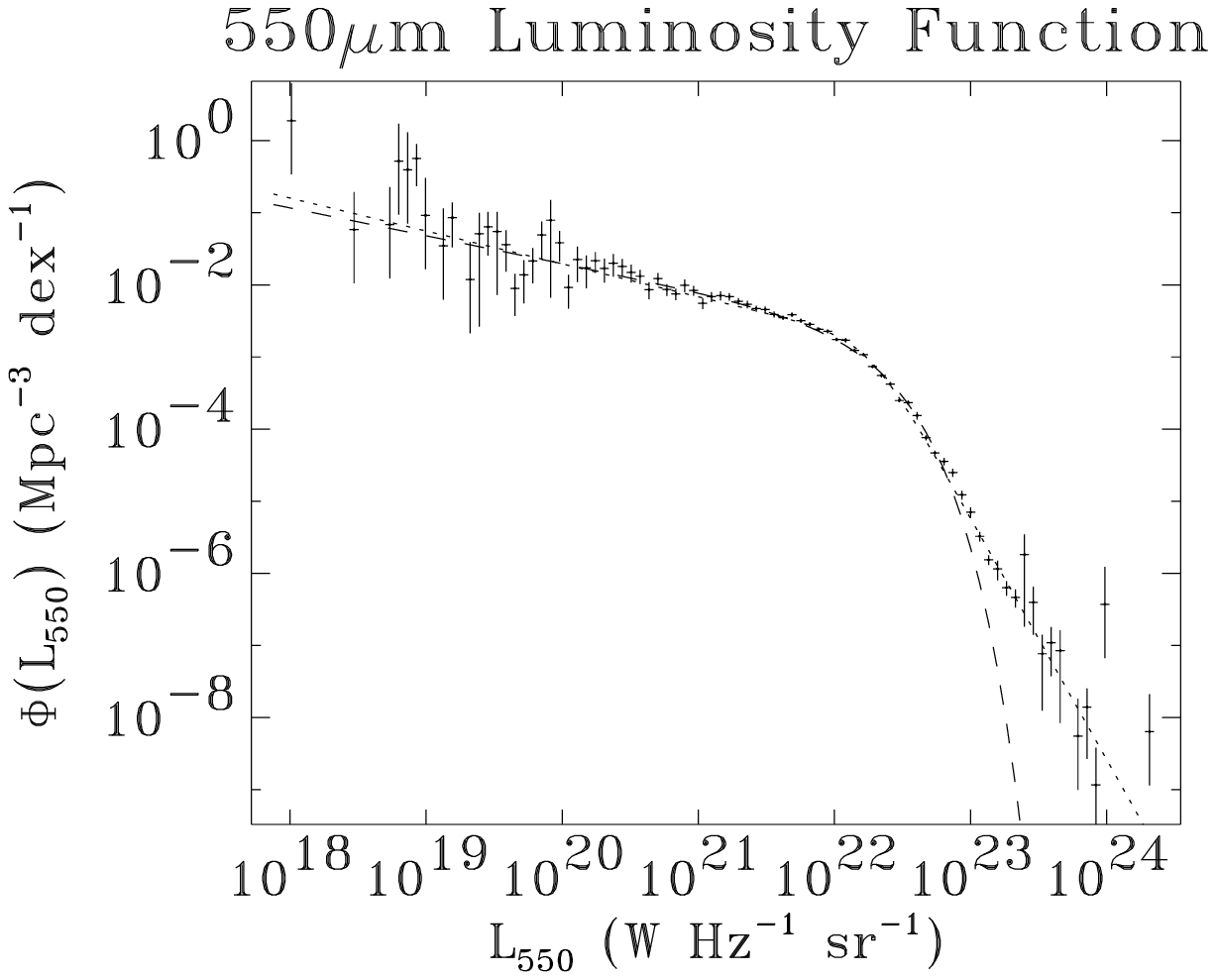}
\caption{\label{fig:lfs}
Local luminosity functions derived from the PSC-z SED ensemble. 
A selection of 
wavelengths relevant 
for ISO, Herschel, Astro-F, BLAST, SCUBA and Planck HFI are shown.
Also plotted are the best fit Schechter and double power
law functions (dashed and dotted lines respectively). 
}
\end{figure*}

\begin{table*}
\begin{tabular}{llllllll}
$\lambda$ & 
$\log_{10}\Phi_{*,S}$ & 
$\log_{10}L_{*,S}$ & $\alpha$ & 
$\log_{10}\Phi_{*,P}$ & 
$\log_{10}L_{*,P}$ & 
$\beta$ & 
$\gamma$\\
($\mu$m) & 
(dex$^{-1}$ Mpc$^{-3}$) & 
(W Hz$^{-1}$ sr$^{-1}$) &  & 
(dex$^{-1}$ Mpc$^{-3}$) & 
(W Hz$^{-1}$ sr$^{-1}$) & 
 & \\
70  & -3.72 & 23.69 & 1.73 & -3.27 & 23.56 & 0.705 & 3.04\\
90  & -3.43 & 23.69 & 1.63 & -2.93 & 23.54 & 0.567 & 2.84\\
160 & -3.09 & 23.55 & 1.50 & -2.91 & 23.61 & 0.558 & 3.47\\
175 & -3.08 & 23.51 & 1.50 & -2.93 & 23.58 & 0.583 & 3.56\\
250 & -2.98 & 23.19 & 1.43 & -2.78 & 23.23 & 0.481 & 3.28\\
350 & -2.93 & 22.80 & 1.39 & -2.92 & 22.96 & 0.536 & 4.28\\
450 & -2.92 & 22.48 & 1.38 & -2.73 & 22.53 & 0.446 & 3.30\\
500 & -2.88 & 22.31 & 1.35 & -2.70 & 22.36 & 0.415 & 3.29\\
550 & -2.91 & 22.19 & 1.38 & -2.74 & 22.25 & 0.456 & 3.35\\
850 & -2.90 & 21.54 & 1.38 & -2.73 & 21.61 & 0.458 & 3.36\\
\end{tabular}
\caption{\label{tab:params}
Best-fit parameters for the luminosity functions shown in
figure \ref{fig:lfs}. The parameters are defined in equations
\ref{eqn:schechter} and \ref{eqn:power}.}
\end{table*}

\begin{figure}
  \ForceWidth{4.0in}
  \hSlide{-1cm}
  \BoxedEPSF{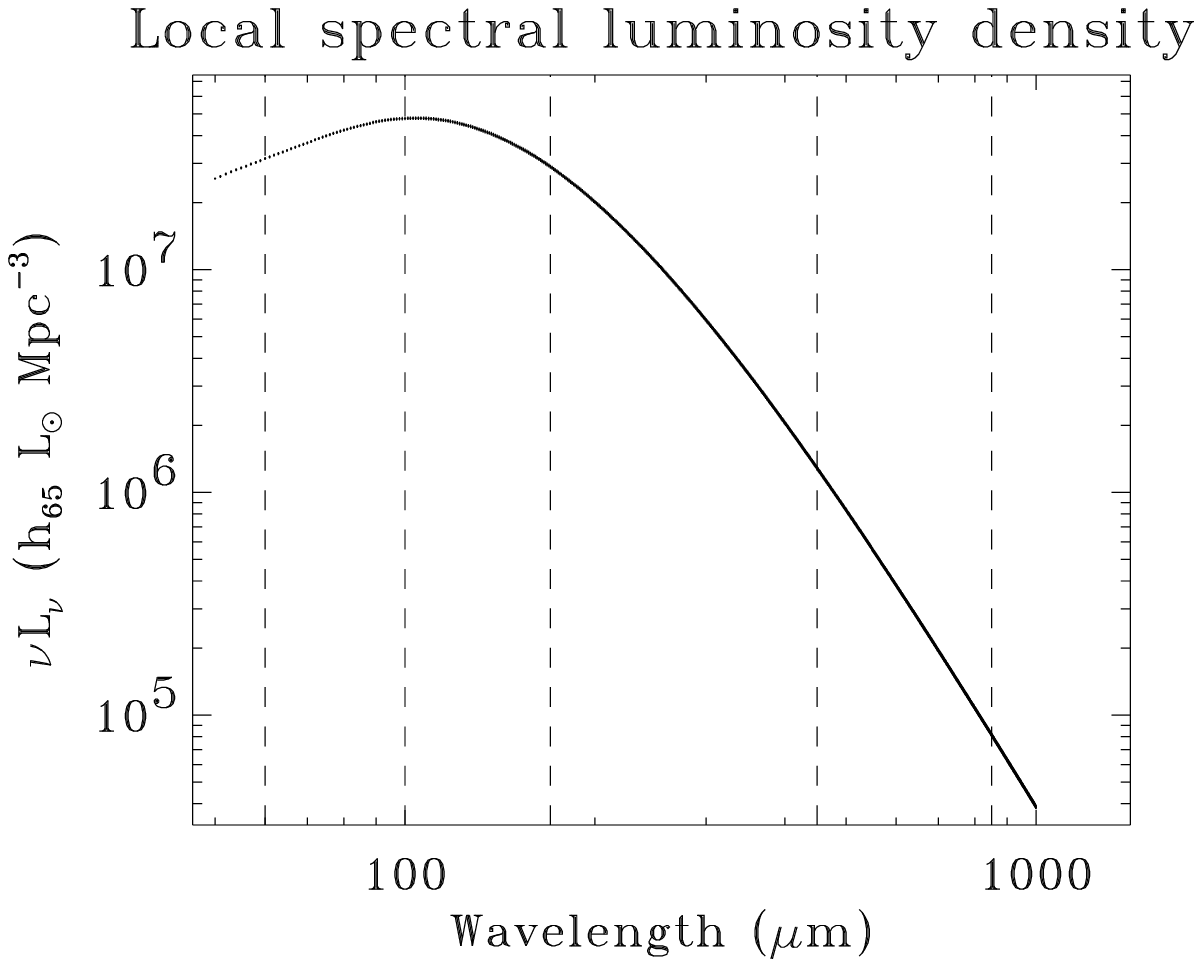}
\caption{\label{fig:luminosity_density}
Local spectral luminosity density. The thickness of the curve
represents the $\pm1\sigma$ error. Vertical lines indicate the
wavelengths where direct estimates are available from colour-colour
relations using PSC-Z, or wavelengths where direct observational tests
verify the methodology. The values of the local luminosity density at 
intervening wavelengths use the two component dust models of PSC-z. 
}
\end{figure}

\begin{figure*}
\vspace*{-0.5cm}
\ForceWidth{3.5in}\hSlide{-3cm}\BoxedEPSF{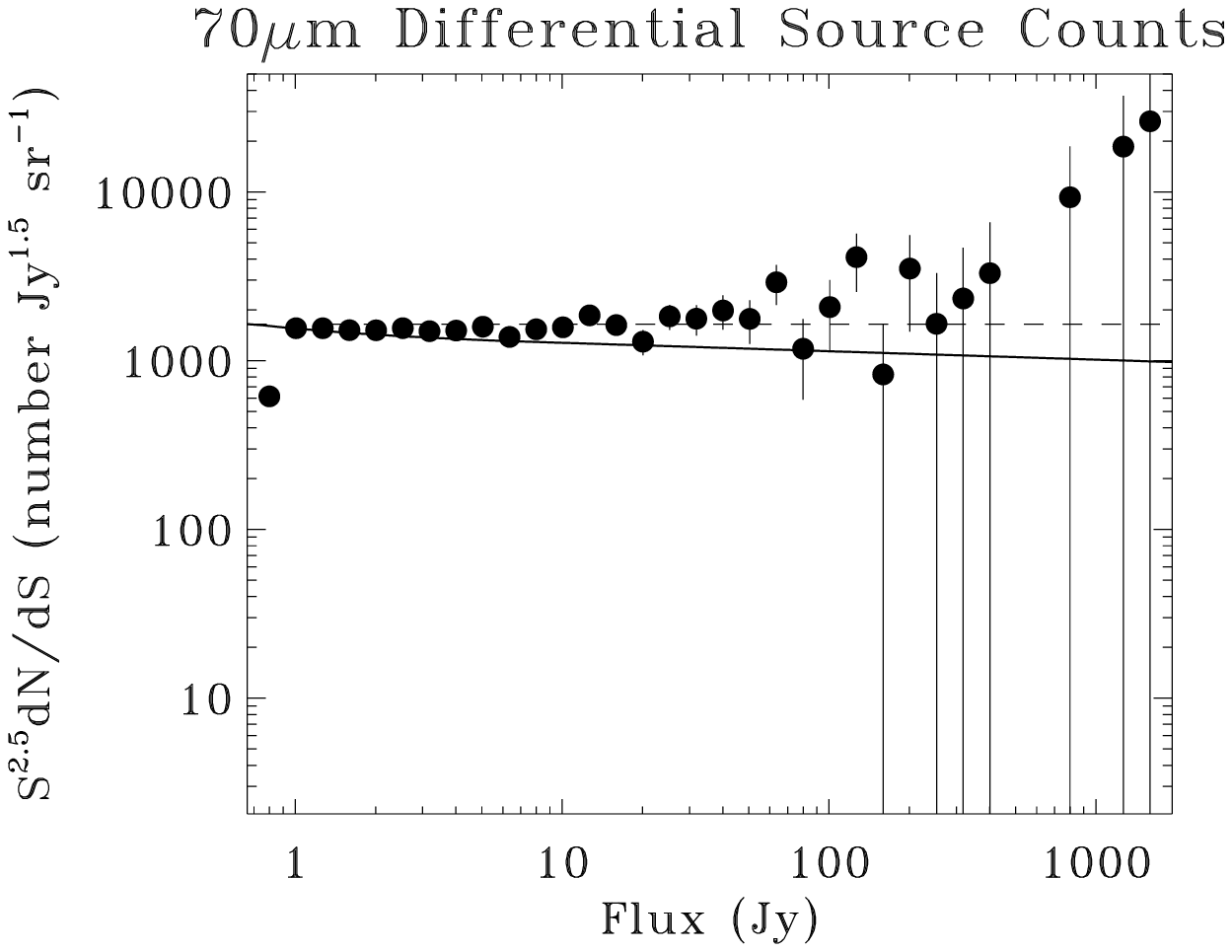}
\ForceWidth{3.5in}\hSlide{5cm}\vspace*{-6.4cm}\BoxedEPSF{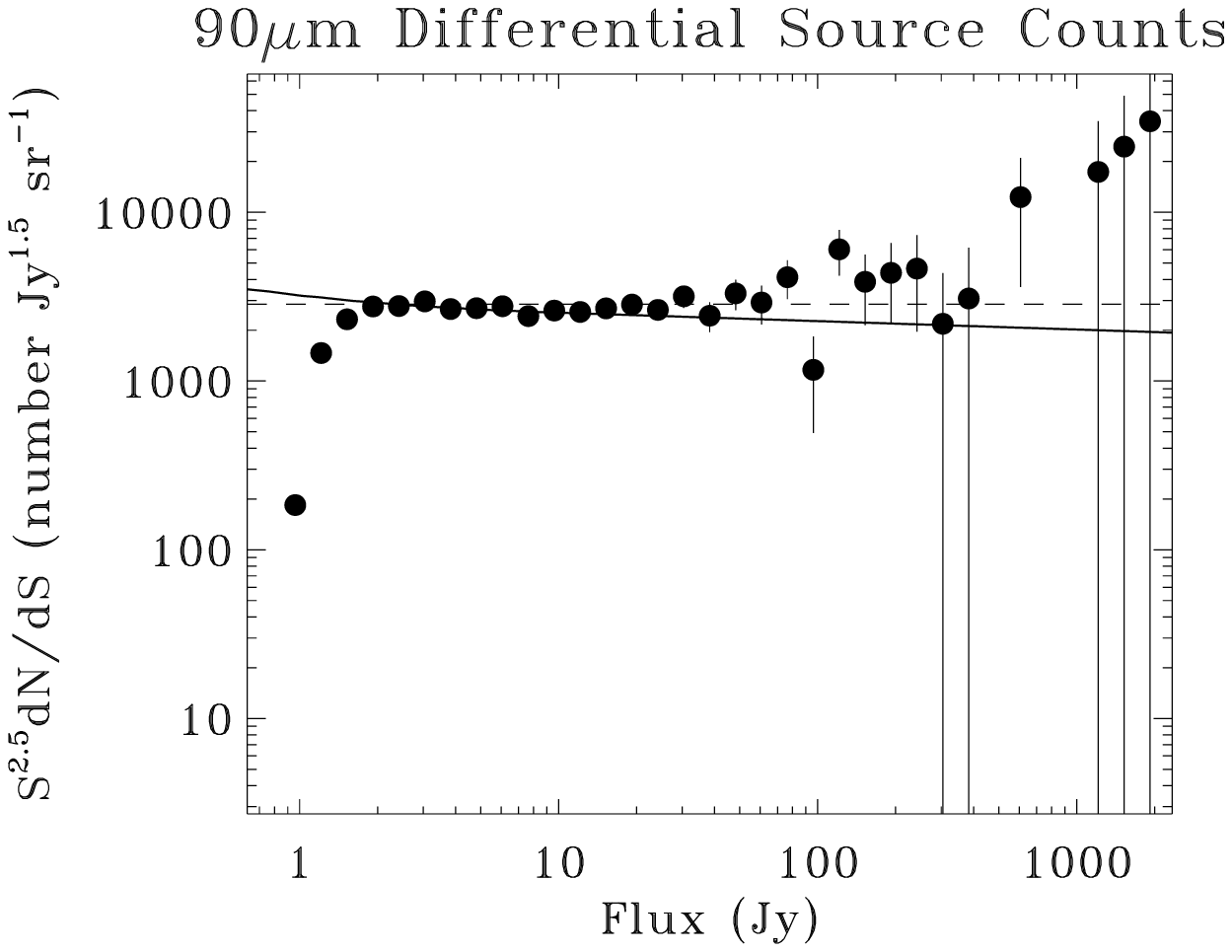}
\vspace*{-1cm}
\ForceWidth{3.5in}\hSlide{-3cm}\BoxedEPSF{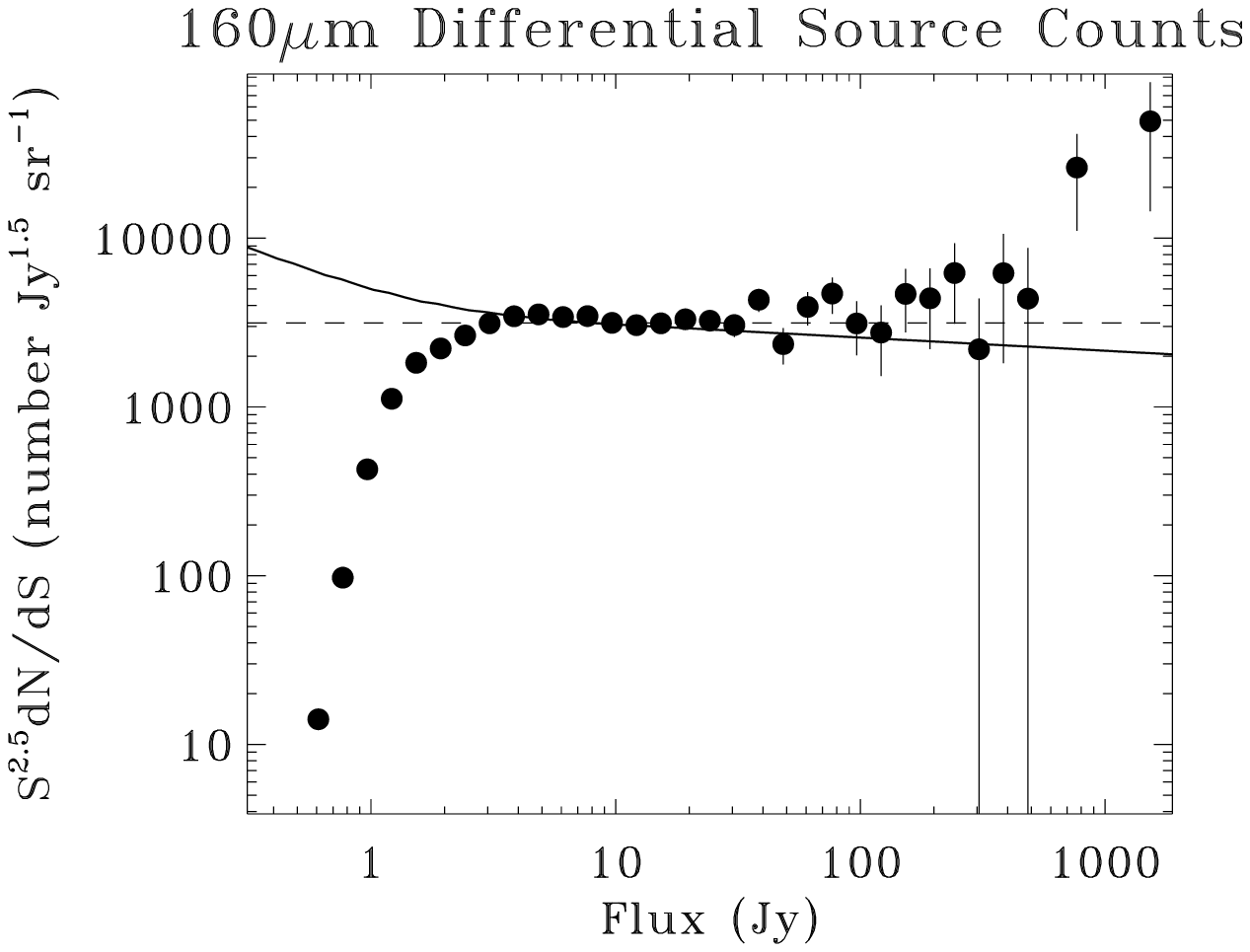}
\ForceWidth{3.5in}\hSlide{5cm}\vspace*{-6.4cm}\BoxedEPSF{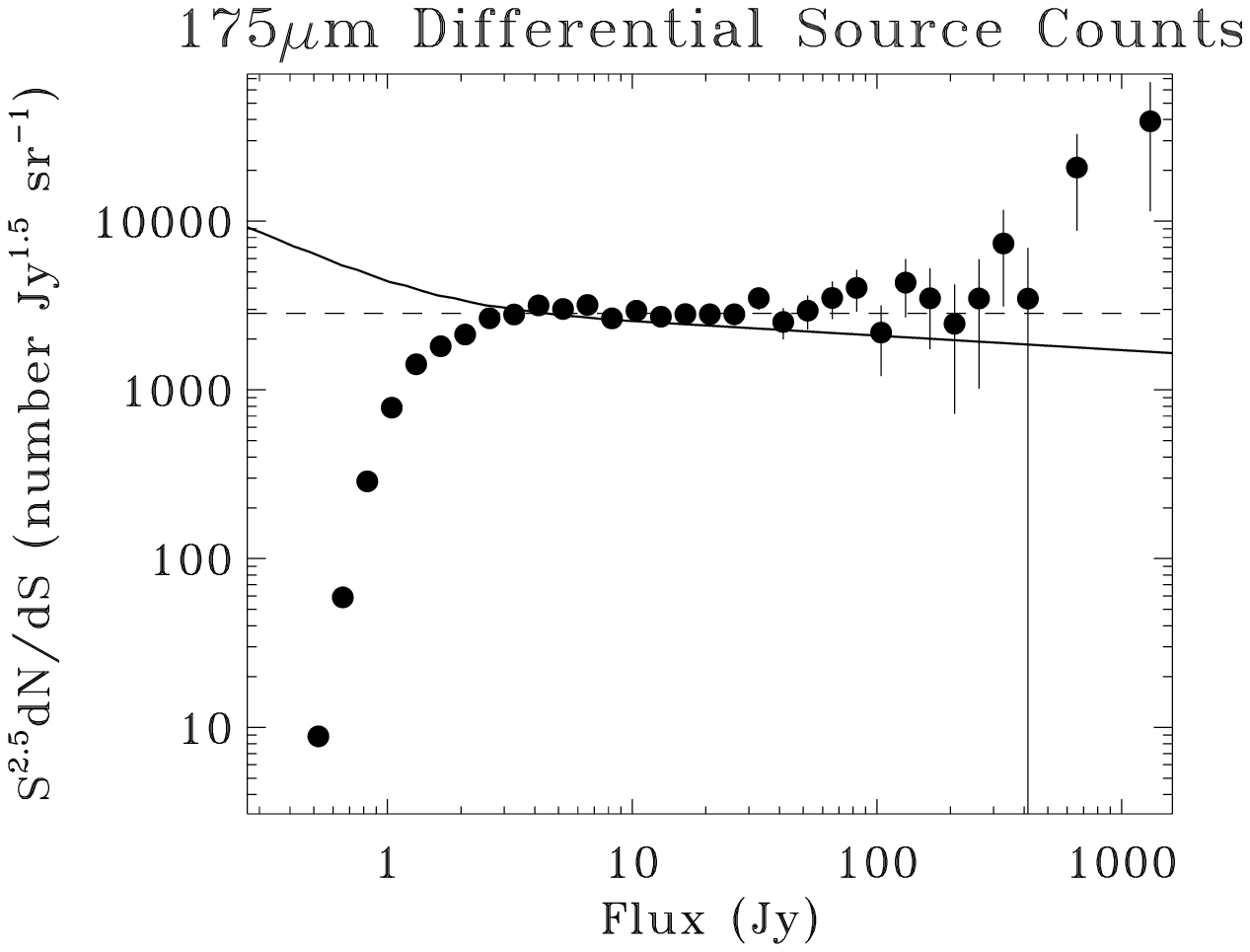}
\vspace*{-1cm}
\ForceWidth{3.5in}\hSlide{-3cm}\BoxedEPSF{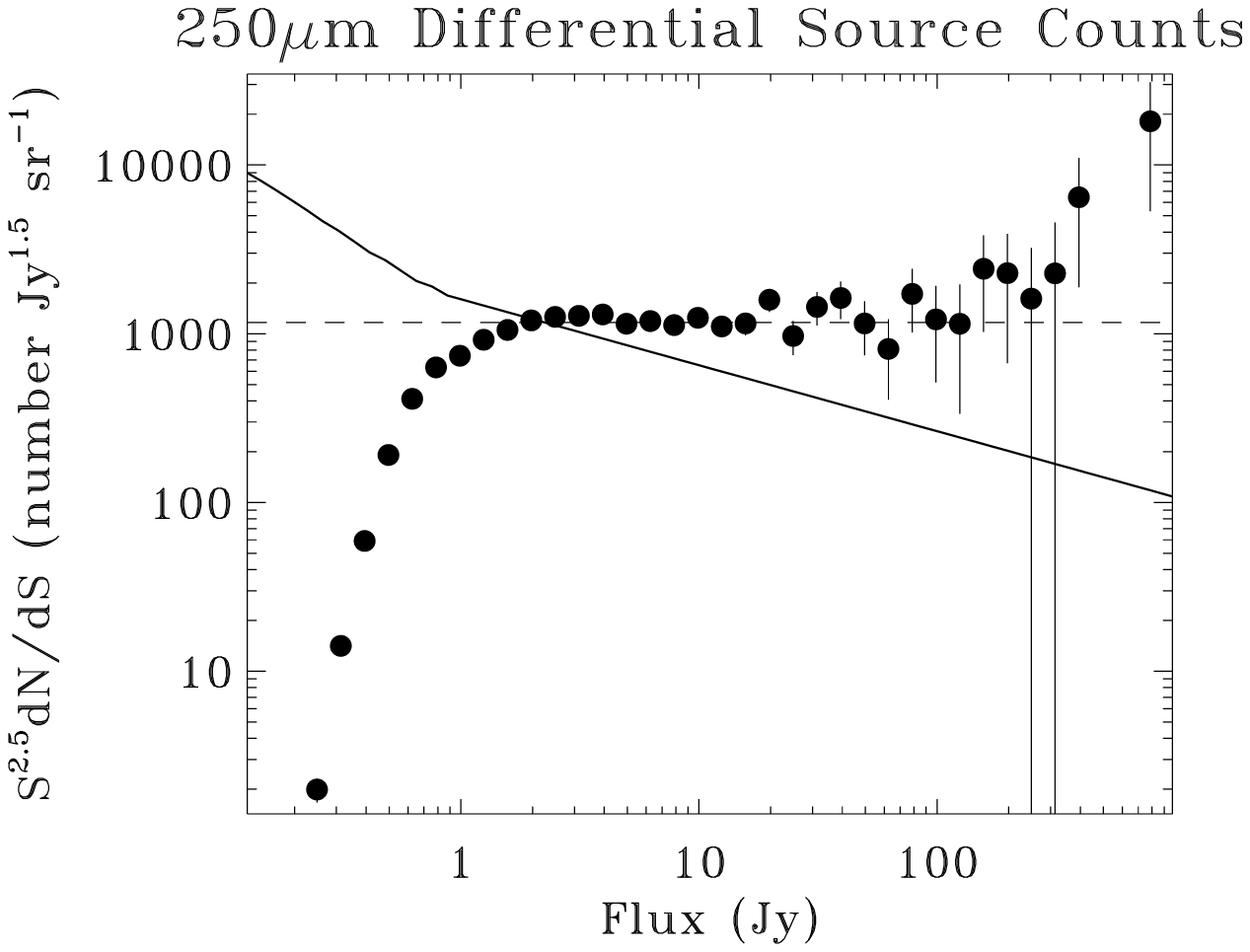}
\ForceWidth{3.5in}\hSlide{5cm}\vspace*{-6.4cm}\BoxedEPSF{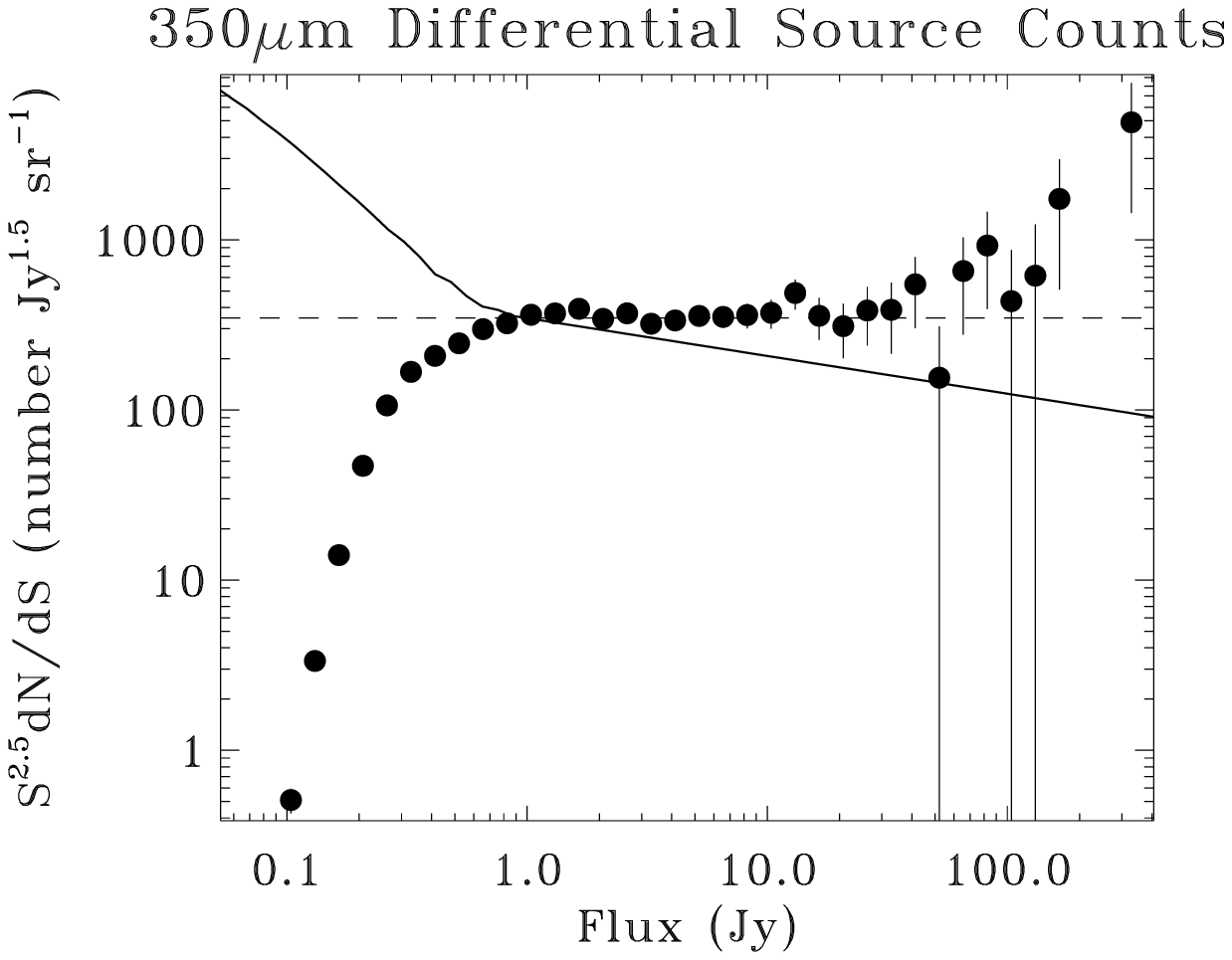}
\caption{\label{fig:counts}
Source counts derived from the PSC-z SED ensemble. Filled symbols show
the source counts of PSC-z galaxies derived in this paper. 
The source count
model of Rowan-Robinson (2001) is overplotted. A selection of 
wavelengths relevant 
for ISO, Herschel, Astro-F, BLAST, SCUBA and Planck HFI are shown. 
Eucludean source counts are horizontal. The asymptotic Euclidean
counts are plotted as horizontal dashed lines and listed in table
\ref{tab:euclidean}. 
The point at which the PSC-Z counts depart from the Euclidean slope at
each wavelength is
a good indicator of the flux level at which sources are missing from
the IRAS PSC-Z survey, compared to a flux-limited survey at the
wavelength in question. The departures from the Euclidean slope at
bright fluxes is due to local large scale structure / cosmic variance
effects. 
}
\end{figure*}
\begin{figure*}
\vspace*{-1cm}
\ForceWidth{3.5in}\hSlide{-3cm}\BoxedEPSF{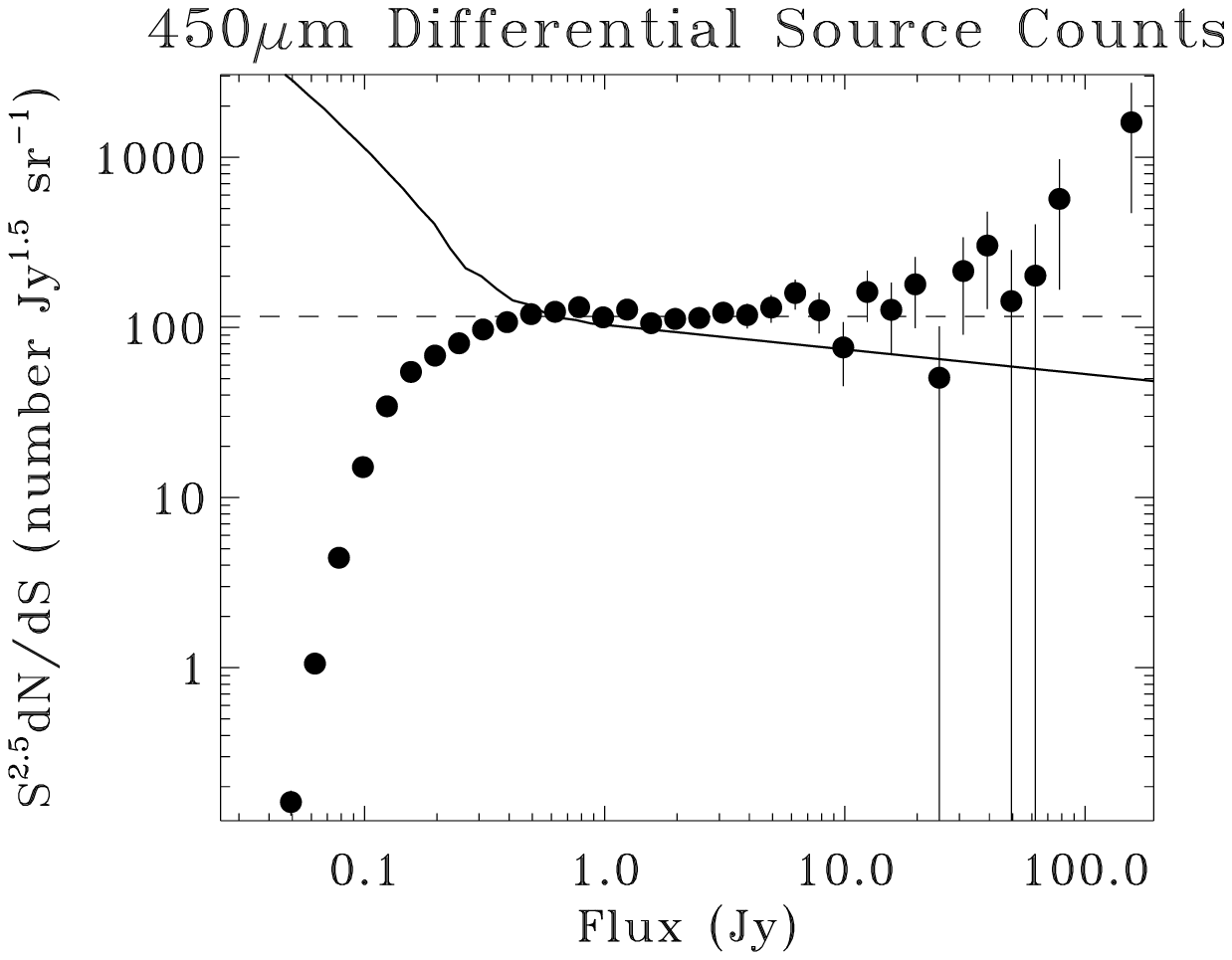}
\ForceWidth{3.5in}\hSlide{5cm}\vspace*{-6.4cm}\BoxedEPSF{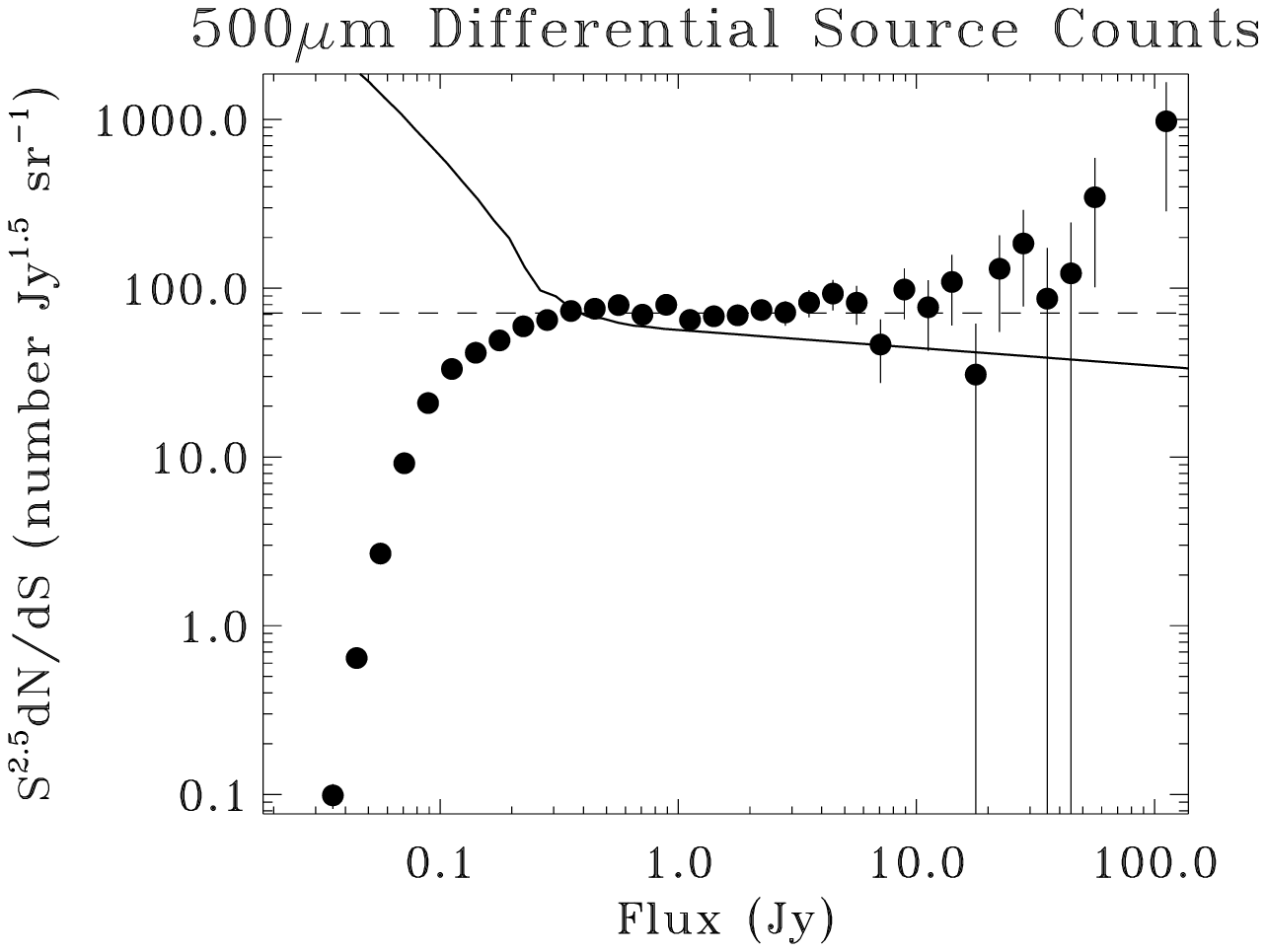}
\vspace*{-1cm}
\ForceWidth{3.5in}\hSlide{-3cm}\BoxedEPSF{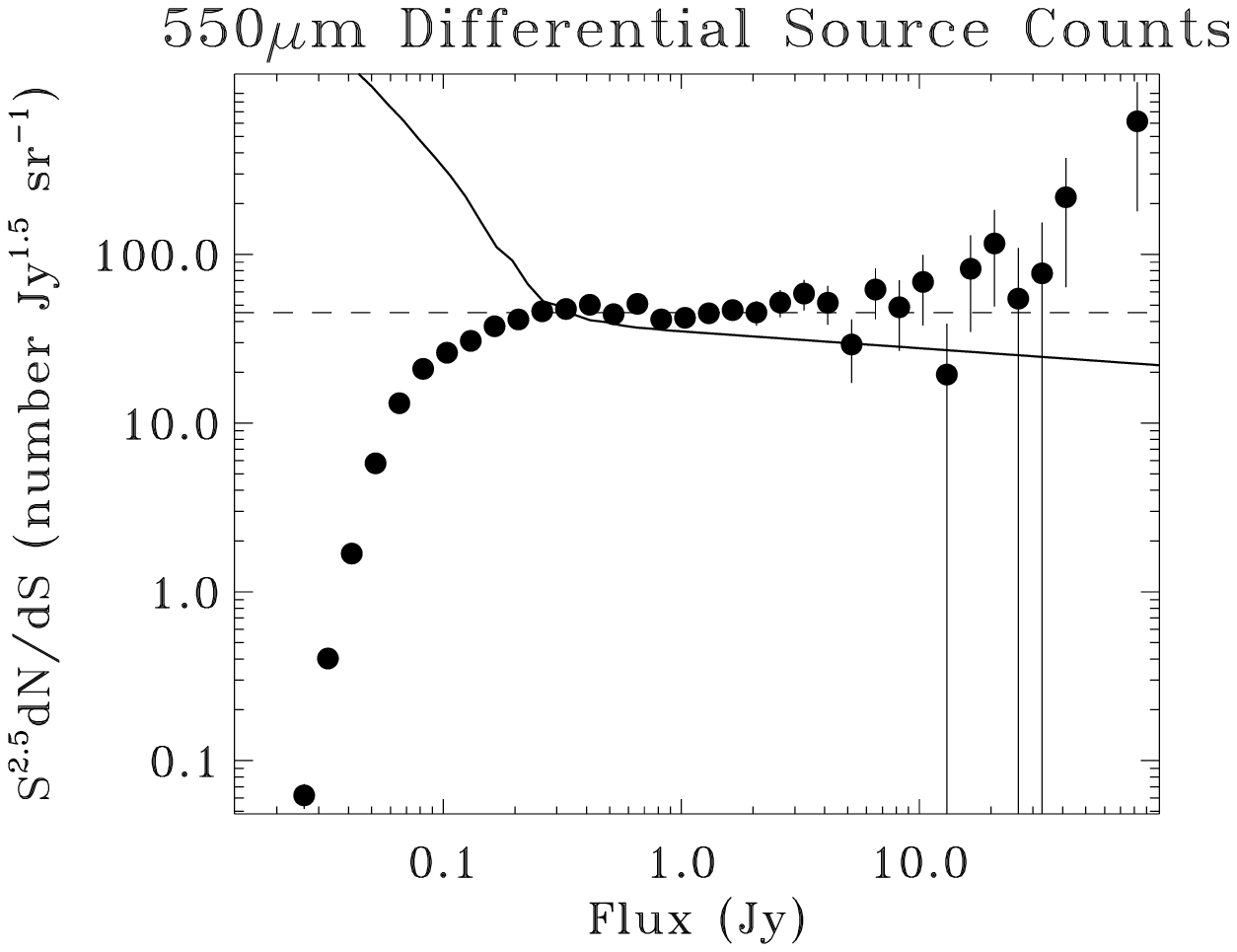}
\ForceWidth{3.5in}\hSlide{5cm}\vspace*{-6.4cm}\BoxedEPSF{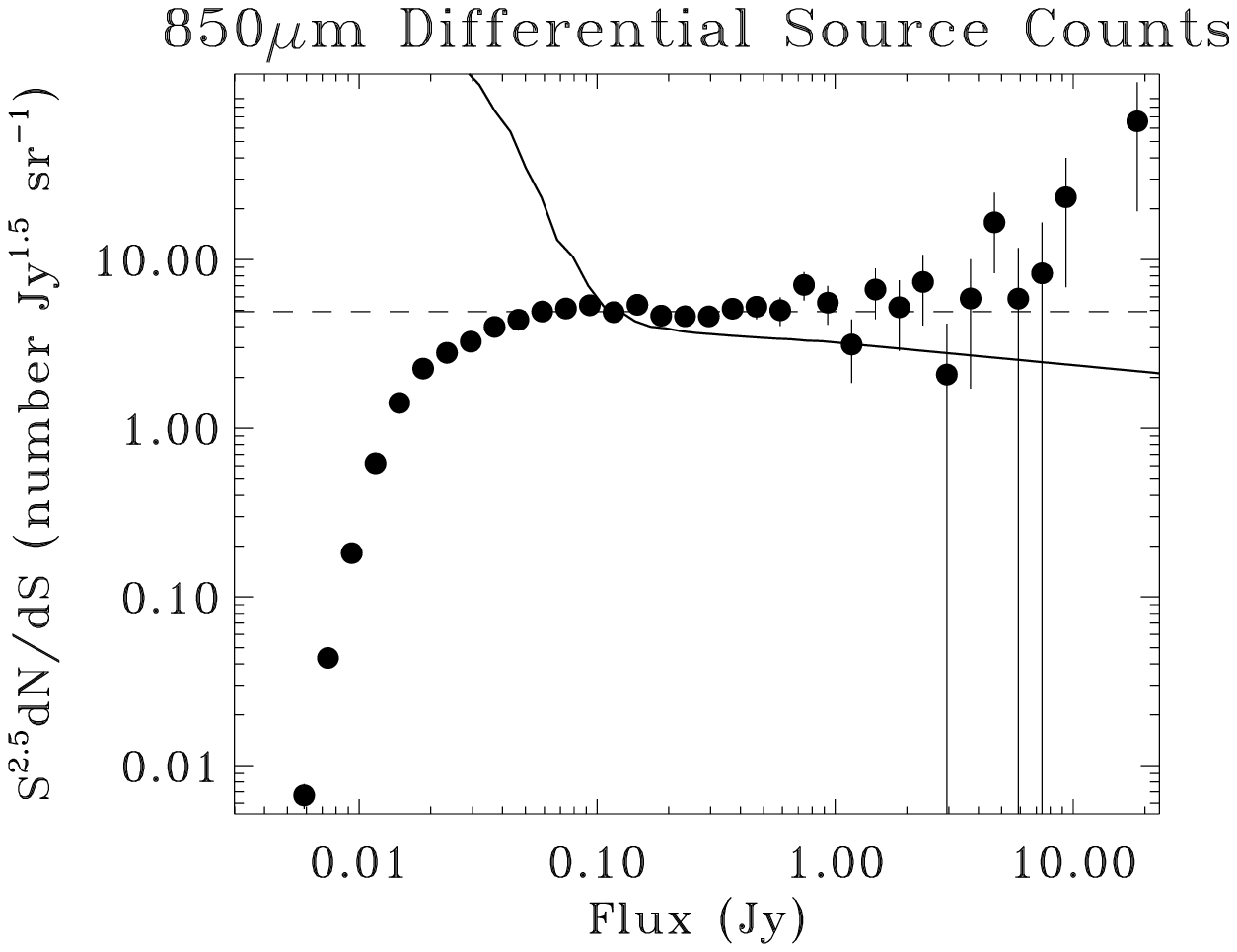}
\caption{\label{fig:counts2}
Continuation of figure \ref{fig:counts}. 
}
\end{figure*}

Unlike the $850\mu$m luminosity function, the $90\mu$m luminosities do
not have direct colour-colour relations to use for their
prediction. We therefore resort to the two-component SEDs to derive
interpolated $90\mu$m fluxes. 
In figure \ref{fig:90_lf} we plot the predicted $90\mu$m luminosity
function for 
PSC-Z, and compare it to the observed luminosity function from the
European Large Area ISO Survey (Serjeant et al. 2001). Pure luminosity 
evolution of $(1+z)^3$ was assumed. 

The $175\mu$m luminosity function cannot be determined directly from
the ISOSS survey, because the completeness of this survey is not
well-understood. In figure \ref{fig:lfs} 
we plot the predicted
$175\mu$m luminosity function from our PSCZ models, and further 
examples of the interpolated luminosity functions possible from our
PSC-z SED ensemble. The $175\mu$m luminosity function is currently 
the best constraint available at around this wavelength before Spitzer
and Astro-F, because the number of
local $175\mu$m-selected galaxies from the European Large Area ISO
Survey (Oliver et al. 2000) and FIRBACK (Dole et al. 2001) is too
small for the construction of a luminosity function. 


Table \ref{tab:params} 
shows the min-$\chi^2$ fit parameters for the local
luminosity functions discussed in this section, using the following
functional forms:
\begin{equation}\label{eqn:schechter}
\Phi = \Phi_{*,S} \ln(10) (L/L_{*,S})^{1-\alpha} e^{-L/L_{*,S}}
\end{equation}
\begin{equation}\label{eqn:power}
\Phi = \Phi_{*,P} / ((L/L_{*,P})^\beta + (L/L_{*,P})^\gamma)
\end{equation}
The Schechter functions give acceptable $\chi^2$ values when
restricted to data less than a few times $L_*$, but in no cases
provide an acceptable fit at the brightest end. The double power law
fit, on the other hand, fits all the available data. 


\subsection{The local extragalactic luminosity
density}\label{sec:luminosity_density} 
The extraordinarily tight constraints on the far-infrared to sub-mm
luminosity functions in section \ref{sec:luminosity_functions}
allow us to make tight constraints on the local
extragalactic spectral luminosity density. This is plotted in figure
\ref{fig:luminosity_density}. The vertical lines indicate where direct
constraints on the PSC-z luminosity densities 
are available via IRAS, SCUBA or ISO. These are spaced roughly in
factors of $2$ in wavelength. Between these wavelengths, the
luminosity densities are based on interpolated flux densities using
our two-component SED models. 
The thickness of the line is the
$\pm1\sigma$ error on the luminosity density.

Remarkably, the local spectral luminosity density itself can be
described to within better than a percent by two grey bodies, despite
being comprised of the sum of $>30000$ individual grey bodies: 
\begin{eqnarray}\label{eqn:luminosity_density}
 & & \hspace*{-0.7cm}I_\nu(0) ({\rm W Hz^{-1} Mpc^{-3}}) = \frac{5.93\times10^{34}}{\lambda^{5.03} (\exp{(705.4/\lambda)}-1)}\nonumber\\
 & & \hspace*{1cm} + \frac{3.87\times10^{31}}{\lambda^{4.50}
(\exp{(318.0/\lambda)}-1)}
\end{eqnarray}
where $\lambda$ is in $\mu$m. and $I$ is in W/Hz/Mpc$^3$. 
This is an instructive lesson: one should not use the two
components of our model (or that of Dunne \& Eales 2001) to argue for
two discrete, physically distinct phases. The two components are a
simple phenomenological model, and do not preclude a continuum of dust
temperatures being present. It is also easy to demonstrate that 
the results of radiative transfer models
of star forming regions (e.g. Efstathiou et al. 2000) 
can often be fit to within a few percent
by a small number of grey bodies, even though a continuum of
temperatures is intrinsic to the models. 

Finally, one can use equation \ref{eqn:luminosity_density} to calculate the
local bolometric luminosity density from thermal dust emission. 
In the interval $30\mu$m-$3$mm, this is found to be 
$2.6\times10^{34}$ 
W Mpc$^{-3}$, i.e. 
$6.4\times10^7 L_\odot$
Mpc$^{-3}$, in very good agreement with the determination by Dwek et
al. 1998. (Differences in the SED ensemble do however generate
differences in our local spectral luminosity density, compared to that
of Dwek et al. 1998.)
The SED models however are not constrained over all this 
range; the local luminosity density over the range $60-1300\mu$m is
$5.3\times10^7 L_\odot$ Mpc$^{-3}$.
Our local bolometric luminosity density differs from that
of Soifer and Neugebauer
(1991) owing to the different bolometric corrections for each galaxy,
but our integrated luminosity density over the range $50-100\mu$m
(i.e. the overlap range in measurements and model validity) is
in good agreement: $4.9\times10^6 L_\odot$ Mpc$^{-3}$, compared to the
result of integrating the Soifer \& Neugebauer data over this range to
obtain $6.3\times 10^6 L_\odot$ Mpc$^{-3}$. The integrated background
in the {\it optical} is much larger: $1.26\times10^8 L_\odot$
Mpc$^{-3}$  (Glazebrook et al. 2003). The fact that the extragalactic
background has a larger luminous energy fraction in the far-infrared
implies stronger relative evolution in the far-infrared luminous
population (e.g. Puget et al. 1996, Hauser et al. 1998, Schlegel et
al. 1998, Rowan-Robinson 2001, Smail et al. 2002).

\subsection{Bright source counts from Astro-F to Herschel}
Figures \ref{fig:counts} and \ref{fig:counts2} 
show the bright source counts at wavelengths
covered by Spitzer, Astro-F, BLAST (Tucker et al. 2004), SCUBA, MAMBO,
Herschel, and Planck.  
The bright slopes are Euclidean as expected (table
\ref{tab:euclidean}), but at faint fluxes the 
counts fall below the Euclidean extrapolation. This is an
incompleteness due to the $60\mu$m flux limit of
PSC-z: of the local galaxies at the faint end of the sub-mm counts,
only the warmer will appear in a $60\mu$m-limited survey. It is
possible to correct
for this using accessible-volume weightings. This is essentially 
equivalent to constructing luminosity functions, discussed above, but
the resulting statistic (confirmation of Euclidean slope) would have 
less utility and interest than the luminosity functions themselves. 
Furthermore, PSC-z is not deep enough to detect the sub-mm blank-field
survey population, both because of the less favourable K-corrections
and because of the low median redshift of PSC-z itself. Therefore,
deviations from the Euclidean slope caused by high-z evolution 
will not be well-determined by
PSC-z. 

\begin{table}
\begin{tabular}{ll}
\hline
Wavelength & $k$\\
($\mu$m)  & (Jy$^{1.5}$ sr$^{-1}$)\\
\hline
$ 70$ & $1642$\\
$ 90$ & $2858$\\
$160$ & $3146$\\
$175$ & $2848$\\
$250$ & $1166$\\
$350$ & $347$\\
$450$ & $116$\\
$500$ & $71.3$\\
$550$ & $45.2$\\
$850$ & $4.92$\\
\hline
\end{tabular}
\caption{\label{tab:euclidean}
Approximate asymptotic Eucludean source count slopes for the 
predicted source counts plotted in figure \ref{fig:counts}. 
These asymptotic scounts follow $dN/dS=kS^{-2.5}$, with the
coefficient $k$ listed in the table as a function of wavelength.
}
\end{table}

How similar are our PSC-Z ``simulated'' catalogues at each wavelength
to flux-limited surveys at those wavelengths? There are two types of
object which could be undetected by IRAS: local galaxies too faint in
the IRAS bands due to their intrinsic luminosities and colour
temperatures, and high-redshift galaxies beyond the upper redshift 
limit of PSC-Z. An
indication of the fluxes at which local galaxies start to be missed is
the point at which the counts depart from the Eucludean slope, in
figures \ref{fig:counts} and \ref{fig:counts2}. Note that this is not
an effect which would alter the luminosity functions, provided there
are no populations which would be missing at {\it any} redshift. 
On the other hand, the level at which high-z populations are seen
depends strongly on the assumed evolution of these galaxies. 
As an indication of the
level of this effect, we overplot predictions from the source
count model of Rowan-Robinson (2001). The slight deviations of the
model from the Euclidean slope is thought to be a numerical artifact
(Rowan-Robinson priv. comm.).

\subsection{High-redshift evolution}
\begin{figure*}
  \ForceWidth{6.0in}
  \hSlide{-3.7cm}
  \BoxedEPSF{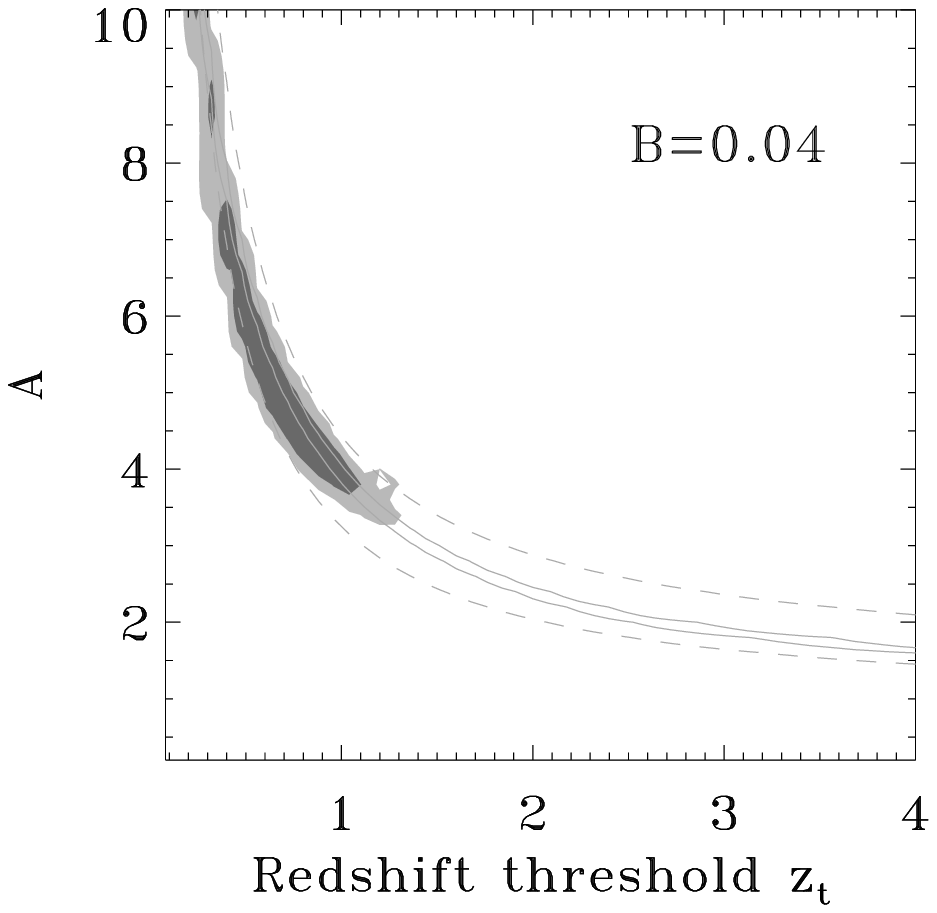}
\vspace*{-10.95cm}
  \ForceWidth{6.0in}
  \hSlide{4.5cm}
  \BoxedEPSF{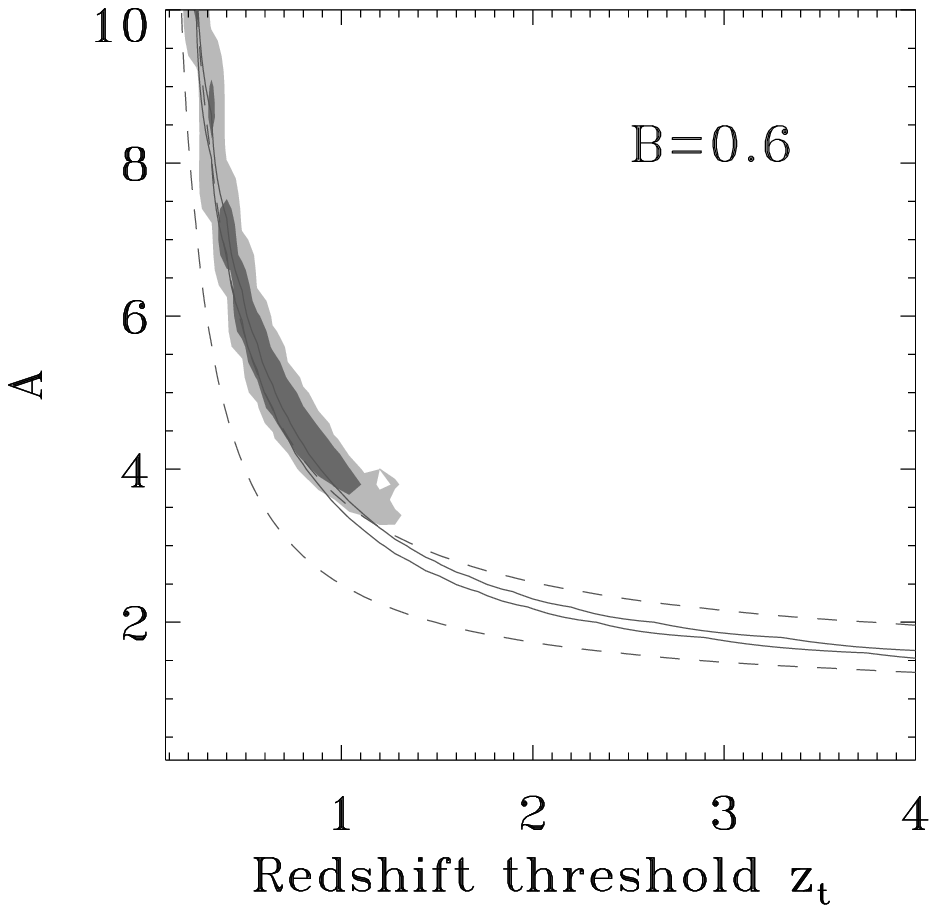}
\vspace*{-2.5cm}
  \ForceWidth{6.0in}
  \hSlide{-3.55cm}
  \BoxedEPSF{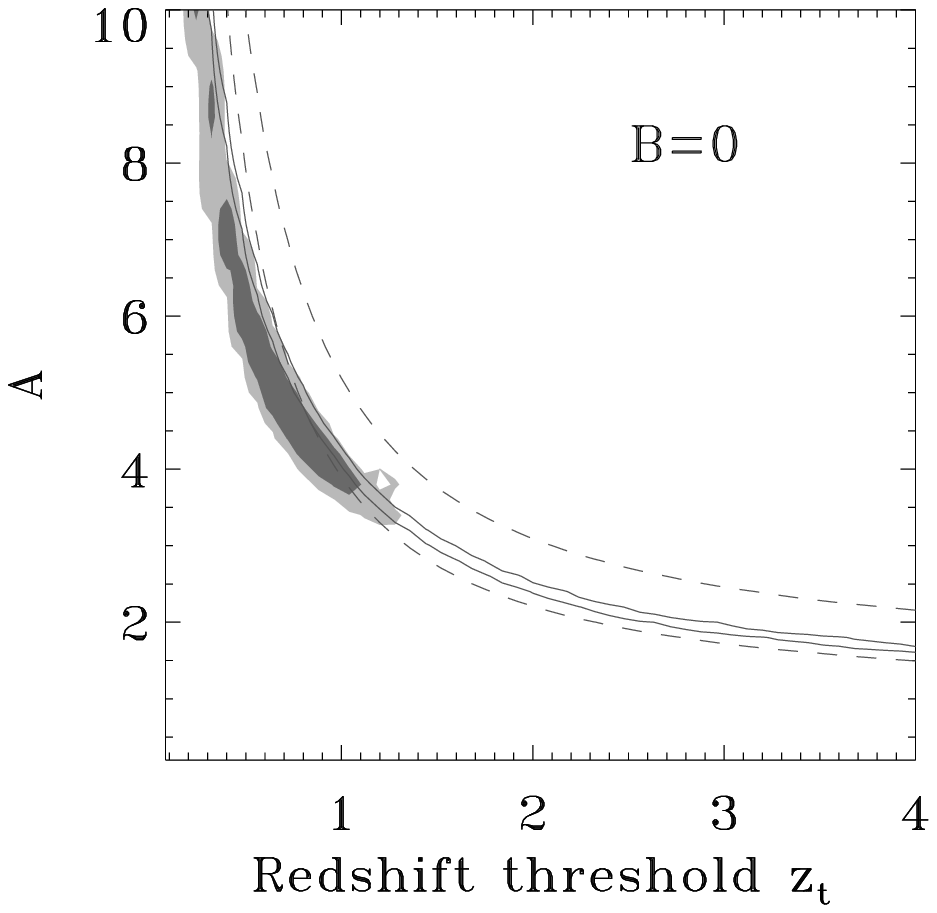}
\vspace*{-10.95cm}
  \ForceWidth{6.0in}
  \hSlide{4.45cm}
  \BoxedEPSF{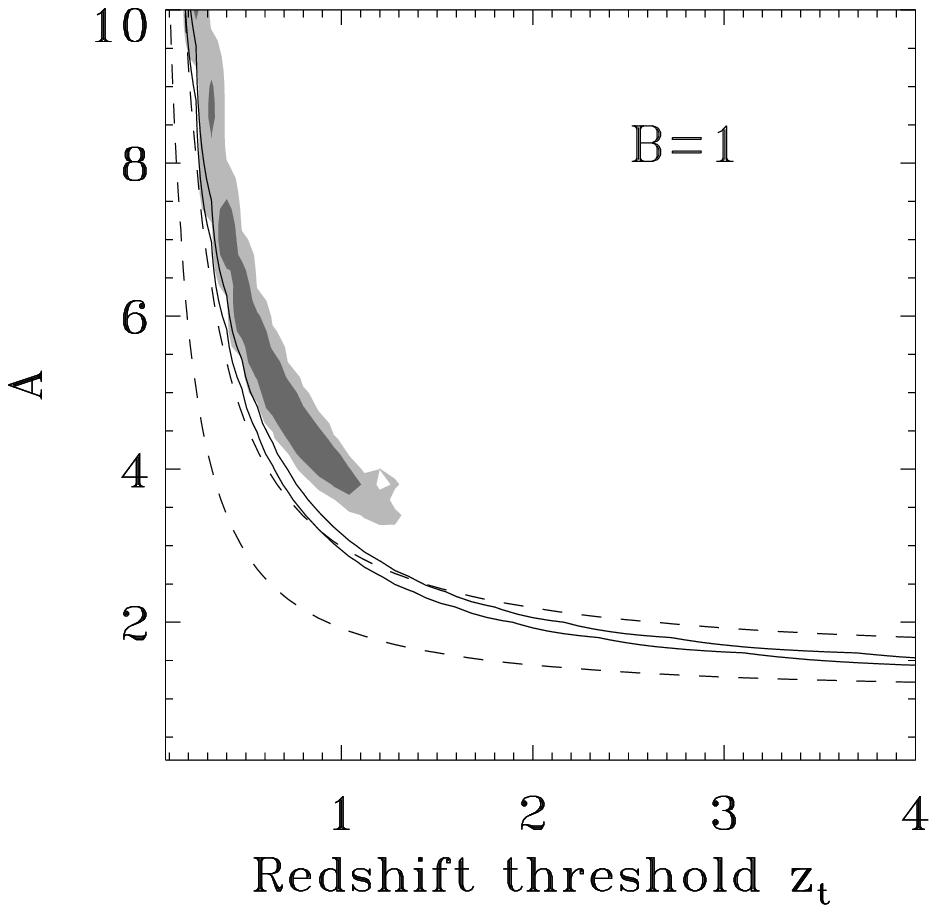}
\vspace*{-0.5cm}
\caption{\label{fig:azt}
Constraints on the parameterisation of the $z\stackrel{<}{_\sim}2$
cosmic star formation history. The definitions of the parameters $A$
and $z_t$ are 
given in equation \ref{eqn:madau_parameterisation}. 
The shaded regions show the $68\%$ and
$95\%$ confidence regions (inner and outer region respectively) for
$A$ and $z_t$, marginalised over $B$. These constraints are derived
from only the extragalactic background light and our 
determination of the local spectral luminosity density. The dashed
lines show the {\it independent} constraints from
$\Omega_\ast=0.003\pm0.0009 h^{-1}$ (e.g. Lanzetta, Yahil, \& Fernandez-Soto
1996), and the full line shows the predicted 
$8$mJy $850\mu$m source count constraint of $N(>S)=320_{-100}^{+80}$
deg$^{-2}$ (Scott et al. 2002) assuming pure luminosity evolution. 
The upper left panel shows the constraints for the $B=0.04$ high-$z$
evolution, which is the maximum likelihood fit from our data (table
\ref{tab:madauparams}). The upper right panel shows the $B=0.6$ model,
which has the highest $f(z>z_t)$ of all models allowed by our
extragalactic background light fits. The $\Omega_\ast$ and $8$mJy
$850\mu$m data are only marginally consistent with this high-$z$
evolution. For comparison, we also plot the $B=0$ case in the lower
left panel, corresponding to $f(z>z_t)=0$, and the $B=1$ case
($f(z>z_t)=$ constant). The $B=1$ case is already excluded by our
extragalactic background fit. 
}
\end{figure*}

Can we use our local luminosity functions (section
\ref{sec:luminosity_functions}) and the local spectral luminosity
density (section \ref{sec:luminosity_density}) to constrain the cosmic
star formation history? 
An exhaustive analysis of the constraints is beyond the scope of this
paper; 
however, 
we note here that the extragalactic background
light is determined entirely by the evolving spectral luminosity
density: 
\begin{equation}\label{eqn:peacock}
I_\nu(\nu) = \frac{1}{4\pi}\int_0^\infty j_\nu (\nu', z) \frac{R_0 dr}{(1+z)}
\end{equation}
where $j_\nu$ is the comoving 
spectral luminosity density, and $R_0 dr$ the
comoving distance element (Peacock 2001; see also Hauser \& Dwek
2001). A very large class of 
evolutionary models including pure luminosity, pure density, and mixed
luminosity/density, satisfy 
\begin{equation}
j_\nu(z)=j_\nu(z=0)f(z)\label{eqn:madau}
\end{equation}
where $f(z)$ is the Madau diagram normalised to $1$ at
$z=0$. We assume the far-infrared luminosity density is proportional
to the volume-average star formation rate, following Rowan-Robinson
2001 (see equation \ref{eqn:omega_star} below). 
A further assumption which underpins this, and indeed almost every
study of the high-$z$ sub-mm population, is that local galaxies can be
found which are reliable templates of the high-$z$ population. This is
not necessarily the case, in which case equation \ref{eqn:madau} will
not hold. Nevertheless, if local galaxies {\it are} reliable templates,
equation \ref{eqn:madau} will hold in very general conditions. 

\begin{figure*}
  \ForceWidth{4.5in}
  \hSlide{-4.0cm}
  \BoxedEPSF{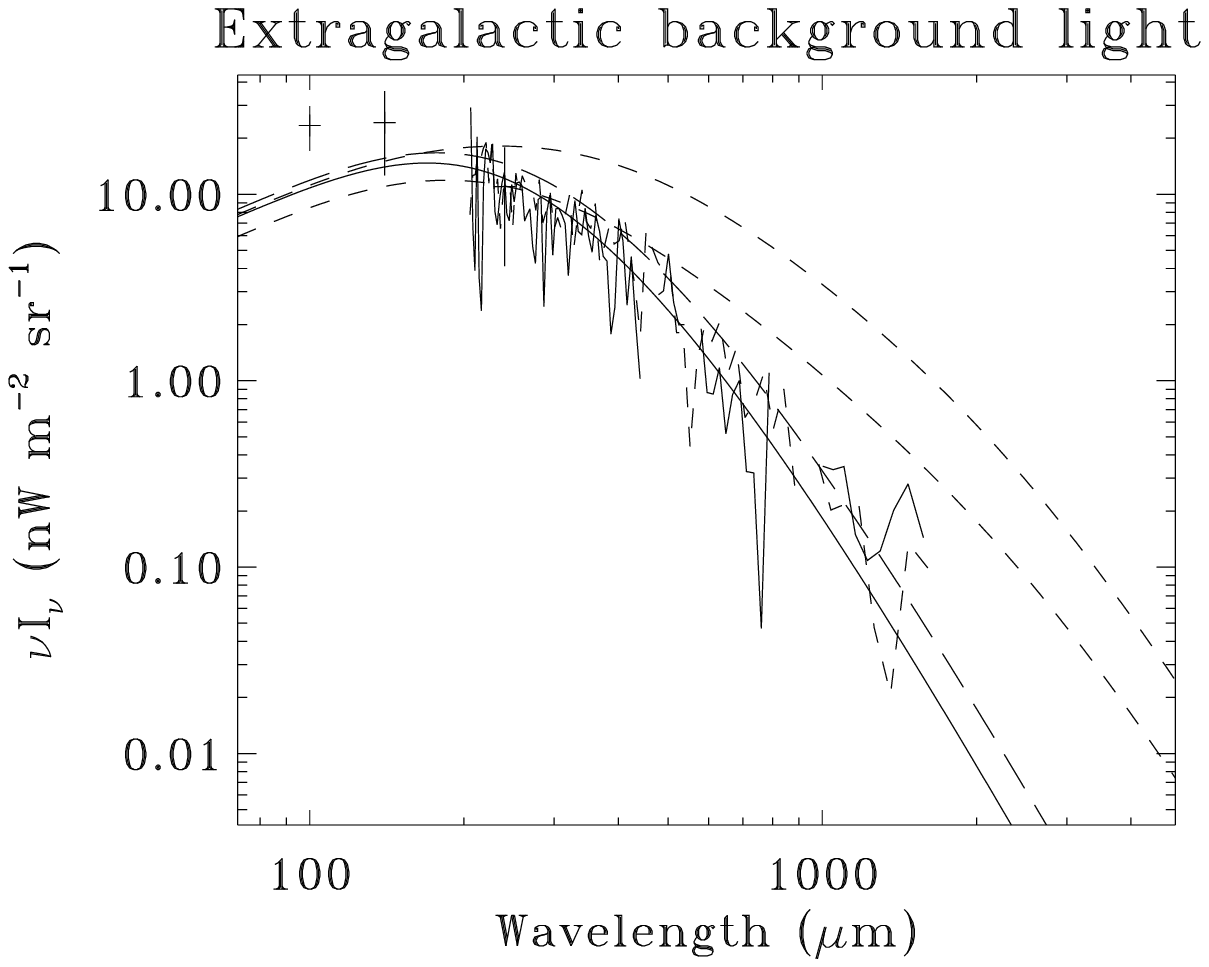}
\vspace*{-8.2cm}
  \ForceWidth{4.5in}
  \hSlide{4.5cm}
  \BoxedEPSF{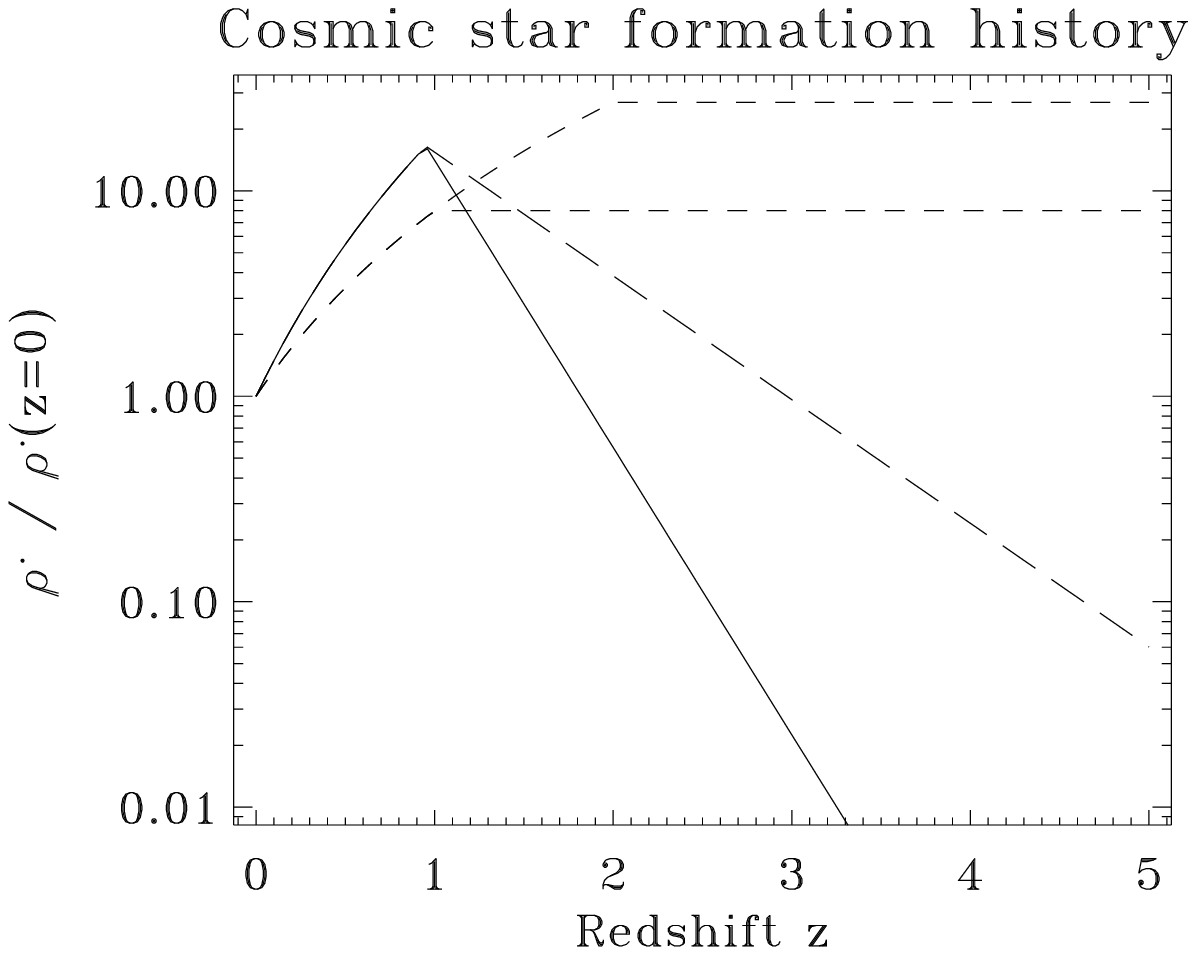}
\caption{\label{fig:ebl}
Left figure shows the 
extragalactic background light (data from Lagache et al. 1999)
modelled by the cosmic star formation history in equation
\ref{eqn:madau_parameterisation}. 
The data longward of $200\mu$m plotted as broken
lines is the FIRAS spectrum: full line is
whole sky, and dashed line is Lockman Hole only. Also plotted are the
DIRBE data points, not included in our fitting. The smooth curves
are models of this spectrum, corresponding to cosmic star
formation histories plotted in the right hand figure, {\it provided
our assumption about the evolving luminosity density (equation
\ref{eqn:madau}) holds.} 
The full line is 
the global maximum likelihood fit quoted in table
\ref{tab:madauparams},  and the long-dashed line has the same
parameters except an enhanced high-$z$ star formation rate ($B=0.25$)
which is marginally ($\sim2\sigma$) inconsistent with the
extragalactic background. The two short-dashed lines demonstrate 
selected alternative models: 
$z_t=1$ and $z_t=2$, both with $A=3$ and $B=1$. 
Note that in general the models with the
higher star formation rates at $z>2$ are also the models predicting the
larger background at wavelengths $\lambda\sim1$mm (including in
particular a comparison of the two short-dashed curves). In general,
models with high volume-averaged star formation rates at $z>2$
overpredict the sub-mm/mm-wave background. 
}
\end{figure*}

We can therefore use the observed extragalactic background light
spectrum to constrain the dust-enshrouded 
cosmic star formation history, {\it independently} of constraints from
redshift surveys of sub-mm blank-field surveys. This has been
attempted by other authors (e.g. Gispert et al. 2000, Dwek et
al. 1998) but without the 
benefit of a robust determination of the local luminosity density
$I_\nu(z=0)$. The authors therefore had to resort to assuming a small
number of template SEDs, and explore the parameter space of $f(z)$
allowed by range of SED models. We have the benefit of a very
well-constrained 
determination of $I_\nu(z=0)$. 

\begin{figure}
  \ForceWidth{4.5in}
  \hSlide{-2.0cm}
  \BoxedEPSF{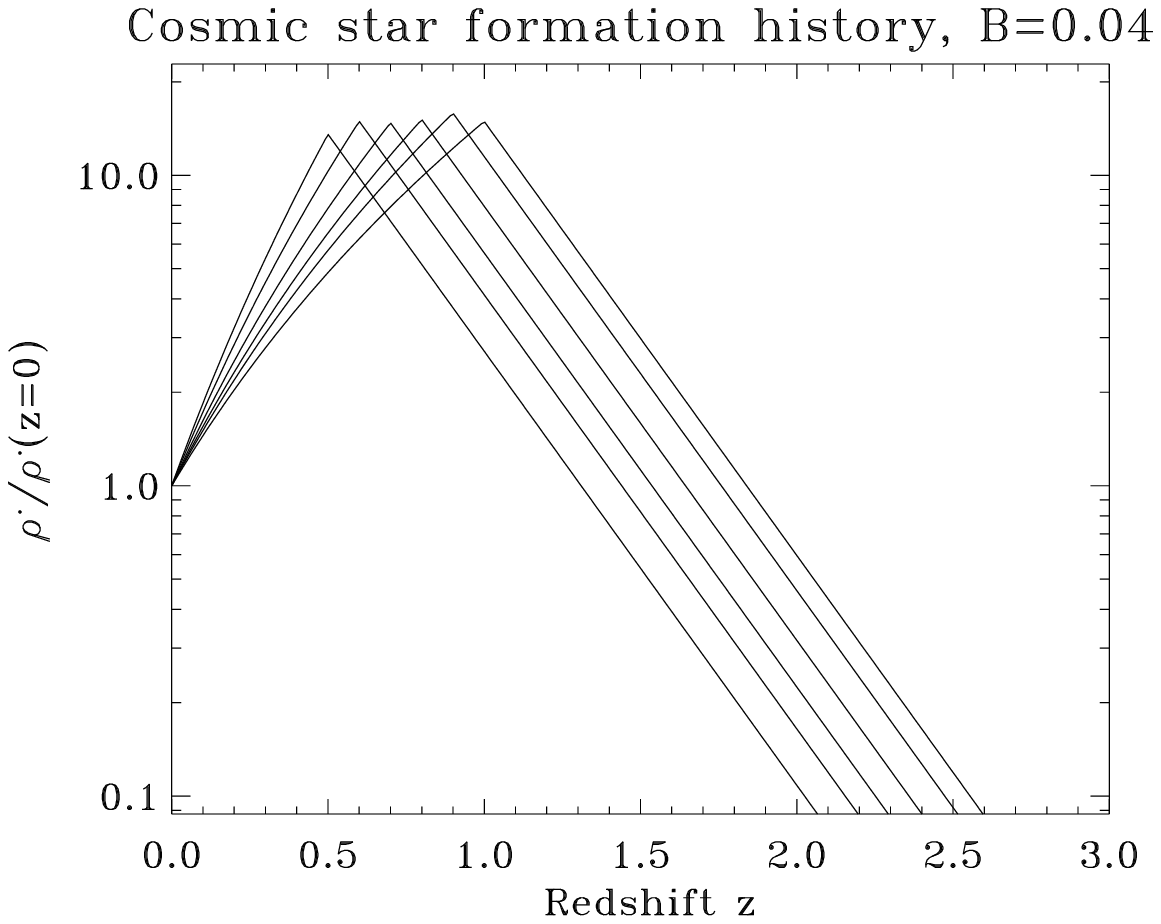}
\caption{\label{fig:extra}
Example of the $\pm1\sigma$ constraints on the cosmic star formation
history defined in figure \ref{fig:azt}. Models are plotted which
satisfy both the sub-mm source counts constrain and the $\pm1\sigma$
likelihood range based on our analysis of the extragalactic background
light. Note that the value of the
peak redshift $z_t$ is {\it strongly} 
correlated with that of the evolution rate
$A$, in that high values of $z_t$ require lower values of $A$, as can
also clearly be seen in figure \ref{fig:azt}. Section
\ref{sec:discussion} discusses how these constraints be resolved with
direct determinations. 
}
\end{figure}

The determinations of the far-infrared
extragalactic background light historically
have fluctuated in the literature, mainly reflecting the 
discovery and differing treatment of systematics (e.g. Puget et
al. 1996, Fixsen et al. 1998). 
We adopt the determination of Lagache et al. (1999), since this is
corroborated by separate analysis of low-cirrus fields, and by the
decomposition using the WHAM H$\alpha$ survey and Leiden/Dwingeloo HI
data (Lagache et al. 2000). We do not fit to DIRBE data shortward of
$200\mu$m, since these predictions rely mainly on an {\it
extrapolation} of our SEDs in their rest-frames rather than an {\it
interpolation}. Warmer dust components are inevitable, regardless of
starburst or AGN activity, but are not incorporated into our SED
models. We defer treatment of these warmer components, their associated
colour-colour diagrams, and of the DIRBE points, to a later paper. 

We found that the extragalactic background light can be fit by a
sequence of judiciously-placed $\delta$-functions in $f(z)$, and that
there is no unique choice for these $\delta$-functions. 
Therefore, a unique 
deconvolution of $f(z)$ from the extragalactic background light is not
possible within the observational errors in the background. 
However, we can still make useful constraints if we restrict
the class of allowed forms for $f(z)$. For simplicity, and to
demonstrate the types of constraints which are available on $f$, 
we model the Madau diagram $f(z)$ with the following simple 
parameterisation:

\begin{eqnarray}\label{eqn:madau_parameterisation}
f(z) = (1+z)^A & [z<z_t] \\ 
f(z) = (1+z_t)^A B^{z-z_t} & [z\ge z_t]\nonumber
\end{eqnarray}
i.e., $(1+z)^A$ power-law evolution to a threshold redshift $z_t$, and
an arbitrary decline thereafter which we have chosen to be a straight
line in the $\log f-z$ plane. 

\begin{table}
\begin{tabular}{llll}
Parameter & Global peak & 68\% & 95\% \\
\hline
$z_t$ & $0.95$ & $0.75^{+0.25}_{-0.05}$ & $0.75^{+0.50}_{-0.15}$\\
$A$   & $4.20$ & $>3.5$ & $>3.3$\\
$B$   & $0.04$ & $<0.52$ & $<0.59$\\
\end{tabular}
\caption{\label{tab:madauparams}
Maximum likelihood parameters for the extragalactic background light
fit. The parameters are defined in equation
\ref{eqn:madau_parameterisation}. The 
$68\%$ and $95\%$ confidence bounds quoted are in each case
marginalised over the other two parameters with an $2<A<5$ top-hat
prior assumed. The global maximum is not affected by the prior on $A$. 
}
\end{table}

The results of this fitting are shown in figure \ref{fig:azt}, where
we plot the confidence limits on $A$ and $z_t$ marginalised over $B$. 
Restricting the parameter space explored to fixed values of $B$
(i.e. instead of marginalising over all $B$ values) favours different
parts of the contoured region: 
low $B$ values favour the low $A$ / high $z_t$ region of the contours,
while high $B$ values favour the high $A$ / low $z_t$ region. 

These constraints on the $A$ and $z_t$ parameter space
are determined only from the comparison of the
extragalactic background light with our tightly-constrained determination of
the local spectral luminosity density, and made no use of (e.g.) the
empirical constraints available on $\Omega_\ast$ or the $850\mu$m
source counts. These latter two constraints are overplotted in figure
\ref{fig:azt}, for comparision with our extragalactic background light
constraint.
The $850\mu$m counts assume pure luminosity evolution to $z=z_t$, in
concordance with galaxy evolution at other wavelengths, and pure
density evolution thereafter. 
$\Omega_\ast$ is determined from the local $60\mu$m
luminosity density $\rho_{60}$ with the following relation
(Rowan-Robinson 2001):  
\begin{equation}\label{eqn:omega_star}
\Omega_* = 10^{-11.13} \xi h^{-2} 
 \frac{\rho_{60}}{1 L_\odot {\rm Mpc}^{-3}}
 \frac{t_0}{10{\rm  GYr}}
\end{equation}
Some source count models of the sub-mm population have 
had difficulty in reproducing this $\Omega_\ast$ constraint 
(e.g. Blain et
al. 1999), though in our case we show below that
there is good agreement with our 
extragalactic background light modelling. 

Figure \ref{fig:azt} can easily be translated into evolution tracks in
the cosmic star formation history, as shown in figure
\ref{fig:extra}. Note however that the parameters in figure
\ref{fig:azt} are correlated, so that the error on the comoving star
formation rate at one redshift is correlated with that at another
redshift, a fact which is rarely recognised (e.g. Gispert et
al. 2000). Figure \ref{fig:azt} is therefore the ``least-misleading''
presentation of our constraints on the cosmic star formation history. 

An alternative to pure luminosity evolution sometimes used in the 
phenomenology of quasar evolution is to assume that the decline at 
the highest redshifts
is pure density evolution, while still treating the increase from zero
redshift as preserving pure luminosity
evolution. We found this made only a
modest difference to our $850\mu$m $8$mJy counts. 
These predictions are not plotted in figure \ref{fig:azt} for clarity,
but in the $B=0.04$ and $B=0.6$ cases, they 
are roughly equivalent to a $1\sigma$ shift in the observed $8$mJy
counts (i.e., a slight shift down and to the left of the $8$mJy counts
constraint). The $B=0$ and $B=1$ cases are non-evolving above the
threshold redshift $z_t$, so there are no changes in the $8$mJy
$850\mu$m counts predictions in these cases. Also, the $\Omega_\ast$
predictions in general 
depend only on the evolution in luminosity density, and not on the
comparative evolution of luminosities and number densities. 

The global peak of the likelihood in this parameter space 
is given in table \ref{tab:madauparams}, as are the $68\%$ and $95\%$
confidence bounds on each of the parameters when marginalised over the
other two. We found acceptable fits with unphysically large values of
$A$ and very small values of $z_t$ (see figure \ref{fig:azt}) so we
adopted an $2<A<5$ top-hat prior in calculating the confidence
bounds. This did not affect the global peak. 
Figure \ref{fig:ebl} shows the global best fit to the extragalactic
background light data. A notable requirement of these fits is a rapid
decline in comoving star formation density above $z_t\simeq 1$. This
is confirmed both by the $\Omega_\ast$ and $850\mu$m counts
constraints, both of which are overpredicted in a $B=1$ model
(i.e. $f(z)=$ constant above $z=z_t$). 

However, this star formation history is immediately 
seen to be in conflict with
determinations of the redshift distribution of sub-millimetre
galaxies (e.g. Chapman et al. 2003) as well as other direct
determinations. How can these be reconciled? We will address this in
the following section.

\begin{figure*}
  \ForceWidth{4.5in}
  \hSlide{-4.0cm}
  \BoxedEPSF{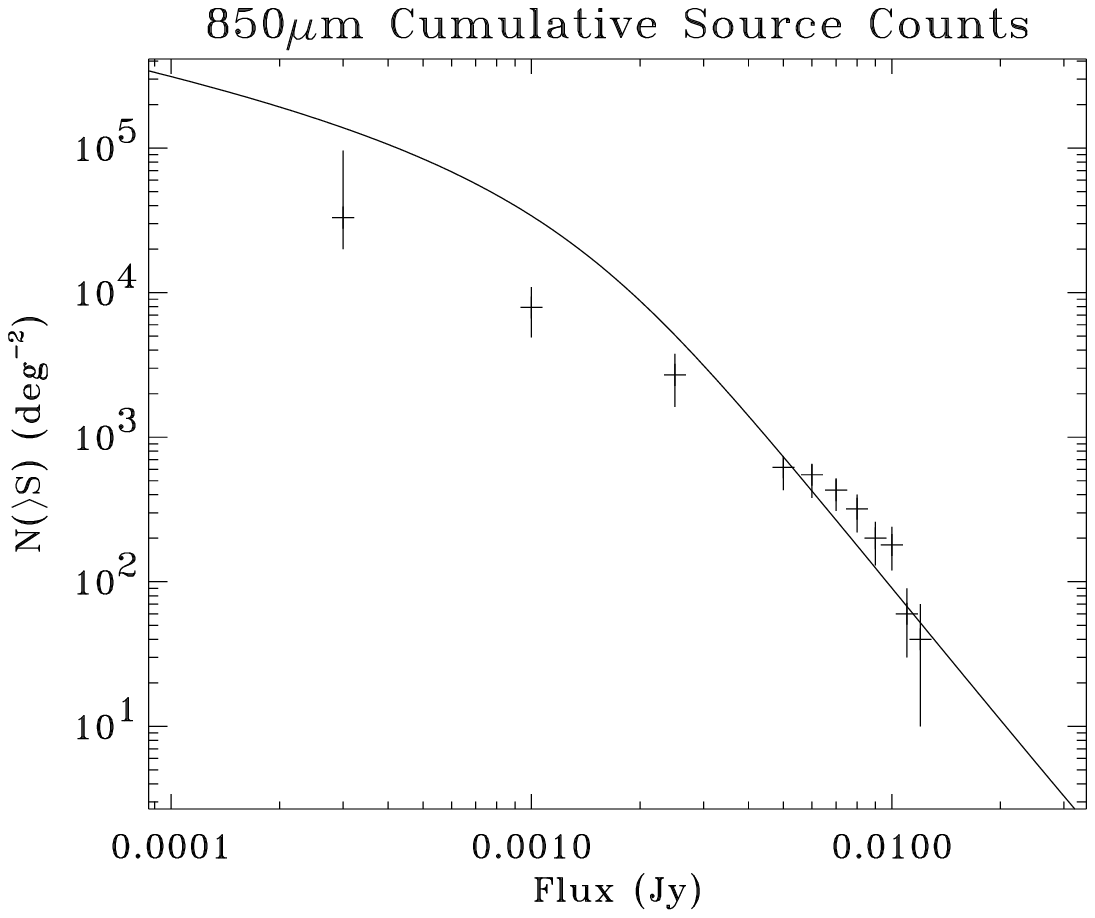}
\vspace*{-8.19cm}
  \ForceWidth{4.5in}
  \hSlide{4.0cm}
  \BoxedEPSF{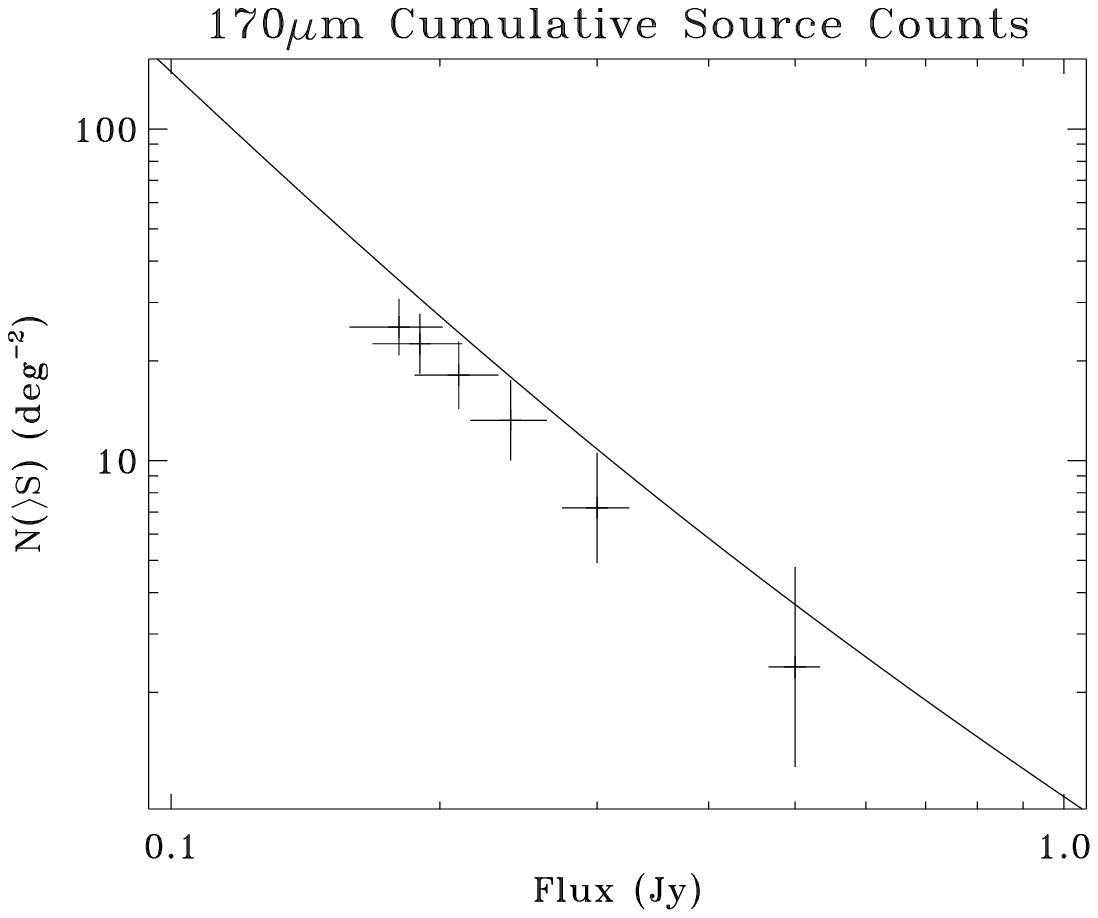}
\vspace*{-0.5cm}
\caption{\label{fig:highz_counts}
Predicted high-redshift 
extragalactic source counts at $170\mu$m and $850\mu$m for
the $A=3.75$, $B=0.04$, $z_{\rm t}=1.0$ pure luminosity evolution
model, compared to the
observed counts from Barger et al. (1999), 
Dole et al. (2001), Scott et al. (2002), Cowie et al. (2002), 
and Webb et al. (2003). Note that {\it none} of these observed counts
were used as constraints in generating the model, except the $850\mu$m
$8$mJy count. In the $850\mu$m plot, each of the data points are from
a different source and (unusually for a cumulative source count plot)
are therefore {\it statistically independent}, with the exception of
the $>8$mJy data which is all from Scott et al. (2002). The
overprediction of the faint $850\mu$m counts is further indication
that the evolution at faint luminosities is less strong that that at
bright luminosities (section \ref{sec:discussion}). 
}
\end{figure*}

\section{Discussion and Conclusions}\label{sec:discussion}

Our determinations of the local far-infrared to sub-mm 
luminosity functions span sufficient luminosity range to identify the
convergence in the local spectral luminosity density. In particular,
the $850\mu$m local luminosity density is found to be
$7.3\pm0.2\times10^{19}$ $h_{65}$ W Hz$^{-1}$ Mpc$^{-3}$ 
solving the ``sub-mm Olbers' Paradox'' (Dunne et al. 2000). 
Our SED predictions rely on the extrapolation of the SLUGS
colour-colour relations, both to higher and lower $60\mu$m
luminosities. Although we have no reason to suspect this extrapolation
is invalid, direct tests with sub-mm photometry of low and high
$60\mu$m galaxies are a key test of our methodology and are an
attractive alternative to large photometric campaigns of local
galaxies for constraining the local sub-mm luminosity functions. 

Using only the observed extragalactic background light and our local
spectral luminosity density, we derive a conditional constraint on the
$z<2$ obscured cosmic star formation history consistent with
$(1+z)^{A}$ evolution with $A>3.5$ to $z\simeq 1$. 
We find that the extragalactic background light, the
$850\mu$m $8$mJy source counts, and the $\Omega_\ast$ constraints all
independently point to a decline in the comoving star formation rate
at $z>1$. However, this is at variance with direct determinations, and
with the constraints on the redshift distribution of sub-millimetre
galaxies (e.g. Chapman et al. 2003). 

Perhaps the most
uncertain of our assumptions is equation \ref{eqn:madau}. This
equation is true in a very general class of cases, including pure
luminosity evolution, pure density evolution and any mixed
luminosity/density evolution models (i.e. those in which the shape of
the luminosity function is invariant with redshift, but not the break
luminosity or the number density at the break). However, if (for
example) the high-luminosity sub-millimetre galaxies evolve more
quickly than their low-luminosity counterparts, then equation
\ref{eqn:madau} will not hold. Stronger evolution at the
high-luminosity end would give the comoving luminosity density a
warmer colour temperature at high redshift, which in turn would permit
higher numbers of sub-millimetre galaxies at high redshift without
violating the extragalactic background light constraint. The warmer
colour temperature may also explain why our models fall short of the
data at $140\mu$m (figure \ref{fig:ebl}), though this may also be due
to our redshifted SEDs being less well-constrained at the shortest
wavelengths.  
The $\Omega_\ast$ constraint would also be alleviated, since the
extrapolation to galaxies with lower luminosities would yield 
less additional star formation. 

A further indication that the evolution may be stronger at high
luminosities is provided by the predicted blank-field high-redshift
source counts at $170\mu$m and $850\mu$m, shown in figure
\ref{fig:highz_counts}. It is important to stress that {\it none} of
the observed source counts in this figure were used as a constraint on
the model, except the $850\mu$m $8$mJy count. 
While the observed 
$170\mu$m and bright $850\mu$m counts are both encouragingly 
well-reproduced by the
model, the faint $850\mu$m source counts are
over-predicted. Suppressing the strength of evolution in the less
luminous population would have the effect of reducing the predicted
$850\mu$m counts at the faint end. 

An alternative solution to alleviating the constraints on
$\Omega_\ast$ is that the initial mass function is more top-heavy at
high redshifts. The infrared luminosities are dominated by
$\stackrel{>}{_\sim}6M_\odot$ stars (Condon 1992), so altering the 
initial mass function can have a strong effect on the $\Omega_\ast$
calculation while leaving the extragalactic background and source
count predictions roughly intact. 
There are good {\it prima facie} reasons for
supposing this might be the case (e.g. Larson 1998 and references
therein), though if it were to entirely explain the discrepancy with
the $850\mu$m redshift distribution, the top heavy initial mass
function would have to exist in low luminosity high-redshift galaxies,
as well as ultraluminous high-redshift galaxies. 

Within the cosmic star
formation histories $f(z)$ we have considered, we can therefore 
conclude that the initial mass function is very different at
high redshift, and/or some differential evolution is {\it required};
all pure luminosity, pure density and mixed luminosity/density
evolution models are excluded by the $850\mu$m blank-field survey
redshift distributions if the initial mass function is the
same at high redshift. Furthermore, comparison with 
the faint $850\mu$m source counts already
indicates the presence of differential evolution.

The issue of mm-wave/sub-mm/radio photometric redshifts is very timely
given the advent of the SHADES survey (van Kampen et al. in prep.,
Mortier, Serjeant et al. in prep.), which will combine SCUBA
and BLAST data with the aim of obtaining photometric redshifts
accurate to $\Delta z\simeq0.5$. The accuracies which are obtainable
are still the subject of debate in the literature (e.g. 
Hughes et al. 2002, Blain et al. 2003) and are limited by the availability of
local templates. It would be interesting to address the photometric
redshift accuracy from the point of view of our SED ensemble. 
However, our results have already indicated that local ultraluminous 
galaxies may not necessarily be templates for the high-z
ultraluminous systems being found by sub-mm blank-field surveys. 
The photometric redshifts will therefore need careful calibration
against spectroscopic redshifts (e.g. Chapman et al. 2003), obtained
for example using Spitzer identifications (e.g. Egami et al. 2004,
Serjeant et al. 2004, Ivison et al. 2004, Frayer et al. 2004,
Charmandaris et al. 2004). 


\section*{Acknowledgements}
We would like to thank the referee, Eli Dwek, for many helpful
suggestions and clarifications to this paper. 
This work was
supported in part by PPARC (grant 
numbers GR/K98728, PPA/G/S/2001/00120, and PPA/V/S/2000/00563) and the
Nuffield Foundation (grant number NAL/00529/G). 
The authors acknowledge the data analysis facilities provided by the
Starlink Project which is run by CCLRC on behalf 
of PPARC.


\begin{thebibliography}{}
\bibitem{} Andreani, P., \& Franceschini, A., 1996 MNRAS 283, 85
\bibitem{} Aretxaga, I., Hughes, D., Gaztanaga, E., Chapin, E.,
Dunlop, J., 2002, preprint astro-ph/0205313
\bibitem{} Barger, A.J., Cowie, L.L., Sanders, D.B., 1999, ApJ 518, L5
\bibitem{} Blain, A.W., Jameson, A., Smail, I., Longair, M.S.,
Kneib, J.-P., Ivison, R.J., 1999 MNRAS 309, 715
\bibitem{} Blain, A.W., Barnard, V.E., Chapman, S.C., 2003 MNRAS 338,
733 
\bibitem{} Chapman, S.C, Helou, G., Lewis, G.F., Dale, D.A., 2003,
preprint astro-ph/0301233
\bibitem{} Charmandaris, V., et al., 2004, ApJS Spitzer Special Issue,
in press 
\bibitem{} Condon, J.J., 1992, ARA\&A, 30, 575
\bibitem{} Cowie, L.L., Barger, A.J., \& Kneib, J.-P., 2002, AJ, 123, 2197
\bibitem{} Dale, D.A., Helou, G., Contursi, A., Silbermann, N.A.,
Kolharar, S., 2001, ApJ 549, 215
\bibitem{} Dale, D.A., Helou, G., 2002, ApJ 576, 159
\bibitem{} Dole, H., et al., 2001, A\&A 372, 364
\bibitem{} Domingue, D.L., Keel, W.C., Ryder, S.D.; White, R.E. III,
1999, AJ 118, 1542
\bibitem{} Dunlop, J., 2001, in Deep millimeter surveys : implications
for galaxy formation and evolution, Proceedings of the UMass/INAOE
conference, University of Massachusetts, Amherst, MA, USA, 19-21 June
2000. Published by Singapore: World Scientific Publishing, 2001. xi,
207 p. Edited by James D. Lowenthal, and David H. Hughes. ISBN:
9810244657, p.11 
\bibitem{} Dunne, L., Eales, S., Edmunds, M., Ivison, R., Alexander,
P., Clements, D.L., 2000 MNRAS 315, 115
\bibitem{} Dunne., L., \& Eales, S.A., 2001 MNRAS 327, 697
\bibitem{} Dwek, E., et al., 1998, ApJ 508, 106
\bibitem{} Eales, S., Lilly, S., Webb, T., Dunne, L., Gear, W.,
Clements, D., Yun, M., 2000 AJ 120, 2244
\bibitem{} Efstathiou, A., \& Rowan-Robinson, M., 2002, MNRAS
submitted 
\bibitem{} Efstathiou, A., Rowan-Robinson, M., Siebenmorgen, R., 2000
MNRAS 313, 734
\bibitem{} Egami, E., et al., 2004, ApJS Spitzer Special Issue, in press
\bibitem{} Frayer, D.T., et al., 2004, ApJS Spitzer Special Issue, in
press 
\bibitem{} Fixsen, D.J., Dwek, E., Mather, J.C., Bennett, C.L.,
Shafer, R.A., 1998 ApJ 508, 123
\bibitem{} Fox, M.J., et al. 2002 MNRAS 331, 839
\bibitem{} Glazebrook, K., et al., 2003, ApJ, 587, 55
\bibitem{} Gispert, R., Lagache, G., Puget, J.-L., 2000, A\&A 360, 1 
\bibitem{} Glenn, J., et al., 1998, SPIE 3357, 326
\bibitem{} Griffin, M.J., Swinyard, B.M., Vigroux, L., 2001, in The
Promise of the Herschel Space Observatory. Eds. G.L. Pilbratt,
J. Cernicharo, A.M. Heras, T. Prusti, \& R. Harris. ESA-SP 460, p. 37 
\bibitem{} Haas, M., Klaas, U., Coulson, I., Thommes, E., Xu, C.,
2000, A\&A 356, L83;  
\bibitem{} Hauser, D.M., et al., 1998, ApJ, 508, 25
\bibitem{} Hauser, D.M., \& Dwek, E., 2001, ARA\&A 39, 249
\bibitem{} Holland, W., et al., 2002, BAAS 201, 121.02
\bibitem{} Hughes, D.H., et al., 2002 MNRAS 335, 871
\bibitem{} Hughes, D., Serjeant, S., Dunlop., J., Rowan-Robinson, M.,
et al., 1998
\bibitem{} Ivison, R.J., Smail, I., Barger, A.J., Kneib, J.-P., 
Blain, A.W., Owen, F.N., Kerr, T.H., Cowie, L.L., 2000, MNRAS 315, 209
\bibitem{} Ivison, R.J., et al., 2004, ApJS Spitzer Special Issue, in
press 
\bibitem{} Kaercher, H.J., \& Baars, J.W., 2000, SPIE 4015, 155
\bibitem{} Kaviani, A., Haehnelt, M.G., Kauffmann, G., 2002, preprint
(astro-ph/0207238) 
\bibitem{} King, A.J., \& Rowan-Robinson, M., 2003, MNRAS 339, 260 
\bibitem{} Lagache, G., Abergel, A., Boulanger, F., D\'{e}sert, F.X.,
Puget, J.-L., 1999, A\&A 344, 322 
\bibitem{} Lagache, G., Haffner, L.M., Reynolds, R.J., Tufte, S.L.,
2000, A\&A 354, 247
\bibitem{} Lagache, G., Dole, H., Puget, J.-L., 2003, MNRAS 338, 555 
\bibitem{} Lamarre, J.M., et al., 2000, ApL\&C 37, 161
\bibitem{} Lanzetta, K. M., Yahil, A., \& Fernandez-Soto, A. 1996,
Nature, 381, 759 
\bibitem{} Larson, R.B., 1998, MNRAS 301, 569
\bibitem{} Lutz, D.,  et al. 2002 A\&A 378, 70
\bibitem{} Madau P., Ferguson H.C., Dickinson M.E., Giavalisco M., Steidel
C.C., Fruchter  A., 1996, MNRAS, 283, 1388
\bibitem{} Oliver, S., et al., 2000, MNRAS 316, 749
\bibitem{} Peacock, J.A., 2000, {\it Cosmological Physics}, CUP,
Cambridge. 
\bibitem{} Pearson, C.P., Matsuhara, H., Onaka, T., Watarai, H.,
Matsumoto, T., 2001, MNRAS 324, 999
\bibitem{} Puget et al. 1996, ApJ 308, L5
\bibitem{} Rowan-Robinson, M.,  2001, ApJ 549, 745
\bibitem{} Saunders, W., et al., 2000 MNRAS 317, 55
\bibitem{} Schlegel, D.J., Finkbeiner, D.P., Davis, M., 1998, ApJ 500,
525
\bibitem{} Schmidt, M., 1968, ApJ 151, 393
\bibitem{} Scott, S.E., et al. 2002 MNRAS 331, 817
\bibitem{} Serjeant, S., et al., 2001, MNRAS, 322, 262
\bibitem{} Serjeant, S., et al., 2003, MNRAS, 344, 887
\bibitem{} Serjeant, S., et al., 2004, ApJS Spitzer Special Issue, in
press 
\bibitem{} Siebenmorgen, R., Kr\"{u}gel, E., Chini, R., 1999, A\&A
351, 495;  
\bibitem{} Smail, I., Ivison, R., Blain, A., Kneib, J.-P., 2002, MNRAS
331, 495
\bibitem{} Shibai, H., 2002, Advances in Space Research, 30, 2089
\bibitem{} Soifer, B.T., Neugebauer, G., AJ, 101, 354
\bibitem{} Stickel., M., Lemke, D., Klaas, U., Beichman, C.A.,
Rowan-Robinson, M., Efstathiou, A., Bogun, S., Kessler, M.F.,
Richter, G., 2000, A\&A 359, 865
\bibitem{} Takeuchi, T.T., Yoshikawa, K., Ishii, T.T., 2003, ApJ, 587,
L89
\bibitem{} Thompson, R.I., Weymann, R.J., Storrie-Lombardi, L.J., 2001
ApJ 546, 649
\bibitem{} Trewhella, M., Davies, J.I., Alton, P.B.,
Bianchi, S., Madore, B.F., 2000 ApJ 543, 153
\bibitem{} Tucker, G.S., et al., 2004, Advances in Space Research, 33,
1793
\bibitem{} Webb, T.M., et al., 2003, ApJ, 587, 41



\end{thebibliography}
\end{document}